\documentclass{cta-author}
{}
{}
{}
\usepackage{algorithmic}
\usepackage{graphicx}
\usepackage{textcomp}
\usepackage{xcolor}
\usepackage{booktabs}
\usepackage{ragged2e}
\usepackage{hyperref}
\usepackage{orcidlink}
\hypersetup{
    colorlinks=true,
    linkcolor=blue,
    citecolor=blue,
    urlcolor=blue
}
\usepackage{stfloats}
\usepackage{physics}
\usepackage{lastpage}
\usepackage{array}
\usepackage{placeins}

\newif\ifhighlight
\highlightfalse
\ifhighlight
  \newcommand{\chg}[1]{{\color{red}#1}}
  \newenvironment{revblock}{\color{red}}{}
\else
  \newcommand{\chg}[1]{#1}
  \newenvironment{revblock}{}{}
\fi

\makeatletter
\def\bstctlcite{\@ifnextchar[{\@bstctlcite}{\@bstctlcite[@auxout]}}
\def\@bstctlcite[#1]#2{\@bsphack\@for\@citeb:=#2\do{%
  \edef\@citeb{\expandafter\@firstofone\@citeb}%
  \if@filesw\immediate\write\csname #1\endcsname{\string\citation{\@citeb}}\fi}\@esphack}
\makeatother

\begin{document}
\bstctlcite{IEEEexample:BSTcontrol}

\supertitle{Research Article}

\title{Non-Maximally Entangled States for Quantum Key Distribution in Underwater Channels: BBM92 Protocol via Kraus Operators}

\author{\au{Nour RIZK$^{1\corr \textsuperscript{\Large\orcidlink{0009-0009-0171-2388}}}$}, \au{Angélique Drémeau$^{1 \textsuperscript{\Large\orcidlink{0000-0001-6325-3695}}}$}, \au{Arnaud Coatanhay$^{1 \textsuperscript{\Large\orcidlink{0000-0003-0101-3105}}}$}}

\address{\add{1}{Lab-STICC, UMR CNRS 6285, ENSTA, Institut Polytechnique de Paris,
29806 Brest Cedex 9, France}
\email{nour.rizk@ensta.fr}}

\begin{abstract}
Underwater optical channels pose significant challenges to the security and reliability of quantum communication systems due to absorption and scattering. In this paper, we investigate the BBM92 entanglement-based quantum key distribution (\chg{EBQKD}) protocol under realistic conditions \chg{over a non-turbulent underwater channel}. Photon pairs are prepared in non-maximally entangled states, and the underwater propagation medium is modeled as a quantum channel \chg{composed of a polarization-independent photon-loss channel and a depolarizing channel}, described within the Kraus operator formalism. The protocol performance is evaluated in terms of \chg{quantum correlations,} quantum bit error rate (QBER) and secret key rate (SKR), analyzed as functions of the entanglement degree and channel degradation parameters. \chg{We derived analytical expressions for the quantum correlations, the QBER and the SKR over the proposed channel model, and validated them through Monte Carlo simulations.} The proposed framework is then applied to various realistic underwater scenarios, considering different water types, namely clear ocean, coastal, and turbid water, as well as varying atmospheric conditions.
\end{abstract}

\maketitle

\section{Introduction}\label{Introduction}

\begin{revblock}
Quantum key distribution (QKD) allows two legitimate parties, Alice and Bob, to establish a secret key by exploiting the fundamental principles of quantum mechanics. Unlike classical cryptographic methods, whose security relies on computational complexity assumptions, QKD can guarantee information-theoretic security, under the security assumptions of the protocol and of its implementation \cite{gisin2002quantum}. Among the developed protocols, BB84 and SARG04 rely on the preparation and measurement of quantum states \cite{brassard1984quantum, scarani2004quantum, rizk:hal-05301294}, whereas other protocols exploit the entanglement between two quantum systems. This is in particular the case of the BBM92 protocol \cite{bennett1992quantum}, in which Alice and Bob perform measurements on entangled photons in order to establish correlations from which a secret key is generated.

The application of these protocols to underwater communications is of particular interest for securing the exchanges between submarines, autonomous underwater vehicles and sensor networks \cite{ji2017towards, zhao2019performance}. However, the aquatic medium imposes strong constraints on photon propagation. Absorption and scattering attenuate the optical signal, while the fluctuations of the medium may introduce additional perturbations \cite{mobley1994light, baykal2022underwater}. Despite these limitations, the blue-green spectral window of seawater makes underwater optical links possible \cite{mobley1994light, kaushal2016underwater}, and the transmission of polarization-entangled photons through seawater has already been demonstrated experimentally \cite{ji2017towards}. These results motivate the study of the performance of QKD protocols under propagation conditions representative of this environment.

Several studies have thus analyzed underwater QKD on the basis of classical propagation and link-budget models \cite{paglierani2023primer, raouf2022performance, ata2025impact}. These approaches quantify the optical losses and their influence on the transmission performance, but they do not directly describe the evolution of the quantum states. This distinction becomes important for entanglement-based protocols, since their performance also depends on the quality of the quantum correlations. Yet, some analyses assume initially maximally entangled states \cite{chalupnik2025realistic}, whereas real sources may produce non-maximally entangled states \cite{gordon2010quantum, white1999nonmaximally, kravtsov2023security}. In addition to this initial imperfection, the interactions with the propagation medium may lead to decoherence \cite{helm2009quantum}. In particular, photon loss and depolarization may affect the performance of the protocol, namely the quantum bit error rate (QBER) and the secret key rate (SKR) \cite{de2020full}. A model accounting for both the initial state of the photons and their evolution in the channel is therefore required.

In our previous works, we studied the BB84, SARG04 and BBM92 protocols over non-turbulent underwater channels with a classical description of the optical link \cite{rizk:hal-05301294, rizk2026eusipco, rizk2026performance}. We then introduced a first approach based on Kraus operators to describe the BBM92 protocol with non-maximally entangled states in random scattering media \cite{rizk:hal-05589560}. Following this approach, the present paper develops a quantum model of the underwater channel that explicitly includes photon loss and depolarization, in order to study their influence on the evolution of the entangled states and on the performance of the protocol.

To this end, the propagation channel is represented by two quantum channels described with Kraus operators, namely a polarization-independent photon-loss channel and a depolarizing channel. We show that these two channels commute, and we derive the analytical expressions of the output density matrix, of the quantum correlations, of the QBER and of the SKR for different degrees of entanglement. The theoretical results are validated by Monte Carlo simulations and then applied to three water types (clear ocean, coastal and turbid) and to different illumination conditions. We also study the influence of the system parameters, namely the degree of entanglement, the depolarization coefficient and the receiver depth, in order to identify the conditions that allow secret key generation and the optimal position of the source along the link.

This study is restricted to a non-turbulent underwater channel, which makes it possible to isolate the effects of optical losses, depolarization and ambient light. Accounting for oceanic turbulence, induced in particular by temperature and salinity fluctuations \cite{paglierani2023primer, raouf2022performance}, is a perspective for the extension of this model.
\end{revblock}

The remainder of this paper is organized as follows. Section~\ref{Section 2} presents the BBM92 protocol and the underwater channel model. Section~\ref{Section3} describes the evolution of the entangled states through the considered quantum channels. Section~\ref{Section4} derives the expressions of the quantum correlations, of the QBER and of the SKR. Section~\ref{Results} presents and discusses the obtained results. Finally, Section~\ref{Section6} concludes the paper and presents the perspectives of this work.

\section{System Model}
\label{Section 2}
\chg{This section describes the entangled source, the geometry of the link and the BBM92 protocol, and then the classical quantities, namely transmittance, background noise and coincidence probabilities, that govern the detection of the photon pairs.}

\begin{figure*}[!t]
    \centering
    \includegraphics[width=0.92\textwidth]{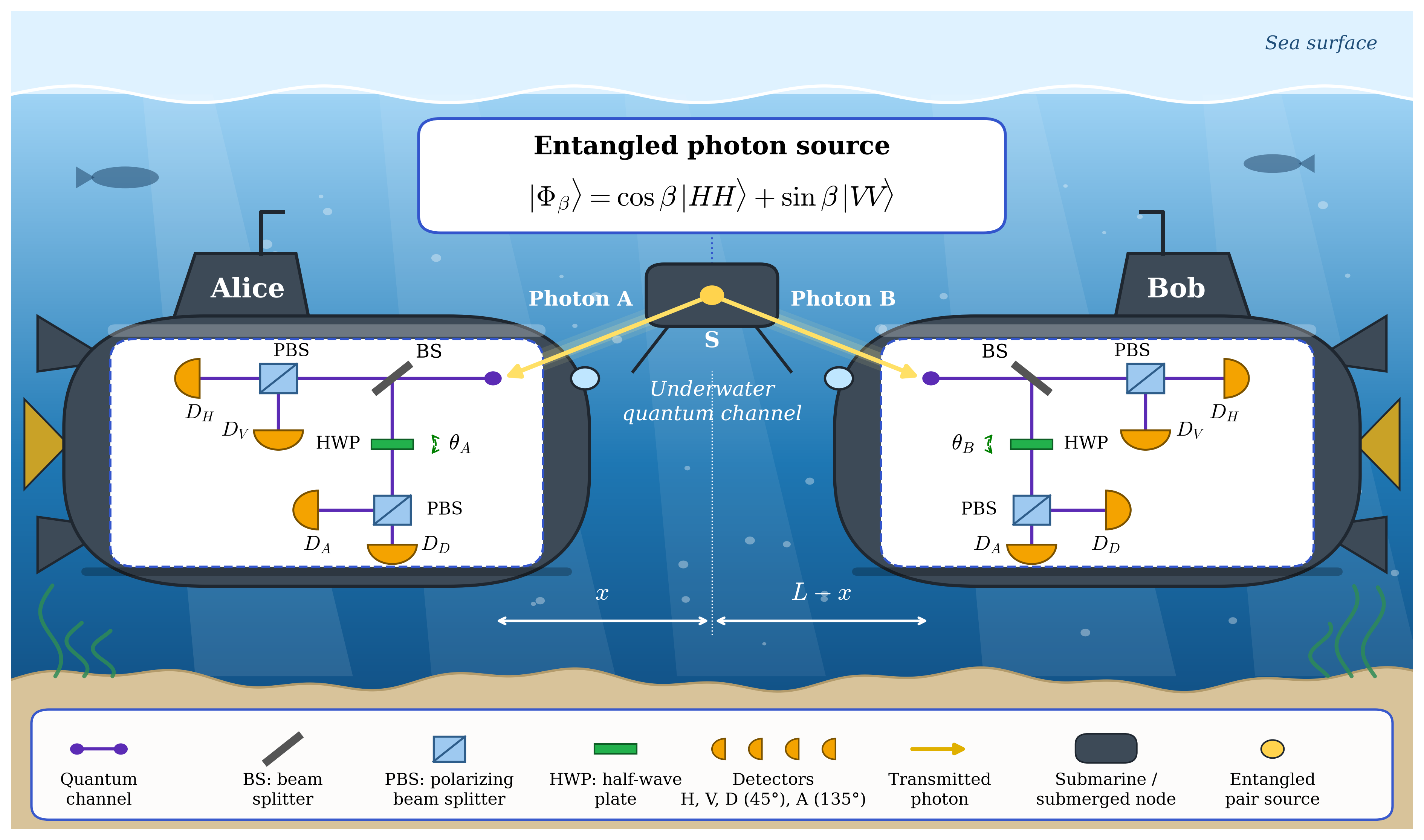}
    \caption{Schematic representation of the BBM92 QKD protocol in an underwater channel\chg{: a submerged source $S$ emits pairs of photons in the state $\ket{\Phi_\beta}$, which propagate over distances $x$ and $L-x$ through the water to the receivers of Alice and Bob, embedded in two submarines}. The orientation $\theta_i$ of each HWP sets the angle of the associated measurement basis.}
    \label{figure_system_model}
\end{figure*}

\subsection{\chg{Non-maximally entangled source}}\label{subsec_source}
In the considered EBQKD system, the entangled photon source generates pairs of polarization-entangled photons prepared in the following pure state
\begin{equation}
\label{non-max-equation}
    \lvert \Phi_{\beta} \rangle = \cos(\beta)\,\lvert HH \rangle
    + \sin(\beta)\,\lvert VV \rangle,
\end{equation}
where \chg{$\beta\in[0,\pi/2]$} is the entanglement
parameter controlling the degree of entanglement between the two photons. \chg{Each photon is measured either in the rectilinear basis $\{\ket{H},\ket{V}\}$ or in the diagonal basis $\{\ket{D},\ket{A}\}$, with $\ket{D}=(\ket{H}+\ket{V})/\sqrt{2}$ and \mbox{$\ket{A}=(\ket{H}-\ket{V})/\sqrt{2}$.} The coefficients $\cos\beta$ and $\sin\beta$ are the Schmidt coefficients of the state, and its degree of entanglement can be quantified by the concurrence
\begin{equation}
\label{eq_concurrence}
\mathcal{C}(\Phi_\beta)=2\,|\cos\beta\sin\beta|=\sin(2\beta).
\end{equation}}
The state reduces to a maximally entangled Bell state \chg{($\mathcal{C}=1$)} when $\beta = \frac{\pi}{4}$, whereas \chg{values of $\beta$ different from $\pi/4$} yield non-maximally entangled states with asymmetric polarization correlations, \chg{ $\beta=0$ and $\beta=\pi/2$ correspond to the separable states $\ket{HH}$ and $\ket{VV}$. Since $\mathcal{C}$ only depends on $\sin(2\beta)$, complementary values such as $\beta=30^\circ$ and $\beta=60^\circ$ lead to the same performance, and the analysis can be restricted to $\beta\le\pi/4$. In this work, $\beta=30^\circ$ ($\mathcal{C}\simeq0.866$) is used as representative non-maximally entangled state.

A more general state of the $\{\ket{HH},\ket{VV}\}$ subspace includes a relative phase $\varphi$, $\cos\beta\ket{HH}+e^{i\varphi}\sin\beta\ket{VV}$. This phase does not change the Schmidt coefficients nor the concurrence, since it can be removed by the local unitary $\ket{H}\bra{H}+e^{-i\varphi}\ket{V}\bra{V}$ applied by Alice, it would however reduce the diagonal-basis correlation by $\cos\varphi$ if left uncompensated. We assume a stable and compensated phase ($\varphi=0$), so that the analysis isolates the effect of the imbalance between the Schmidt coefficients.} The degree of entanglement directly influences the strength of the quantum correlations shared between Alice and Bob, and therefore plays a central role in the protocol's performance, as will be shown in the subsequent analysis.

\subsection{\chg{Link geometry and BBM92 protocol}}
As illustrated in Fig.~\ref{figure_system_model}, one photon from each entangled pair is directed to Alice, located at a distance $x$ from the source, while the other is sent to Bob, located at a distance $L - x$, where \mbox{$L$ denotes} the total link length\chg{, so that the two arms have lengths $\ell_A=x$ and $\ell_B=L-x$}. Both photons propagate through independent underwater quantum channels, which introduce photon loss and polarization degradation due to absorption, scattering, and depolarizing effects inherent to the aquatic medium. These channel impairments act on the entangled state and progressively degrade the quantum correlations, whose modeling is detailed in Section~\ref{Section3}.

In contrast to BB84 protocol \cite{brassard1984quantum}, where Alice prepares and sends single photons to Bob, the BBM92 protocol introduced by Bennett, Brassard, and Mermin in 1992 relies on correlation measurements between pairs of entangled photons sent to Alice and Bob \cite{bennett1992quantum}. Each photon pair is described in a four dimensional Hilbert space defined by the basis states $\{\ket{HH}, \ket{HV}, \ket{VH}, \ket{VV}\}$. In the ideal version of BBM92,
the source \chg{produces} a maximally entangled Bell state such as
$\ket{\Phi^{+}} = (\ket{HH} + \ket{VV})/\sqrt{2}$. In this work, we generalize this model and consider the non-maximally entangled state \(\ket{\Phi_{\beta}}\) of Eq.~\eqref{non-max-equation}, so that $\beta$ becomes an additional degree of freedom for characterizing the behavior of BBM92 under different channel conditions.

Upon reception, Alice and Bob independently and randomly select one of the two measurement bases \chg{defined in Section~\ref{subsec_source}}, the rectilinear basis or the diagonal
basis, to encode bits\mbox{ 0 and 1.} Following the quantum transmission, they communicate over an authenticated classical channel to compare their chosen bases, discarding all results where the bases do not match. The retained results constitute the raw key, which contains residual errors introduced by channel noise and source imperfections. These errors are estimated through the QBER, defined as
\begin{equation}
\label{QBER_BBM92_def}
\text{QBER}_{\text{BBM92}} =
\frac{\text{number of erroneous correlated bits}}
     {\text{total number of correlated bits}},
\end{equation}
and subsequently corrected via classical error-correction protocols, after which privacy amplification is applied to extract a final secret key whose length is quantified by the SKR. It is well established that the BBM92 protocol is considered secure only as long as the QBER remains below the analytical limit of 11$\%$ \cite{waks2002security}. Beyond this value, errors arising from channel noise and from the non-maximal entanglement of the state can no longer be clearly distinguished from those introduced by an eavesdropper (Eve), and no secret key can be securely extracted.

The position $x$ of the source along the link offers an additional degree of freedom in the system model. Symmetric placement \mbox{($x=L/2$)} results in equal loss and noise conditions for both users, while asymmetric placement introduces unbalanced channel impairments that directly affect the observed correlations and the achievable SKR. \chg{The influence of $x$ is quantified in Section~\ref{sec_position}.}

\subsection{Impact of the Underwater Channel on Quantum Transmission}
\label{Classical channel}
During propagation, some photons are lost, and the polarization of the remaining photons can be changed by random scattering in the channel. Because of these effects, the polarization measured by Alice and Bob can be different from the state prepared at the source, which weakens their correlations. \chg{In this section, the link budget is described by classical quantities, namely the transmittance and the background noise, whereas the action of the medium on the two-photon state is described in Section~\ref{Section3}.}

\begin{enumerate}
    \item \textbf{\chg{Attenuation and transmittance}}
\begin{revblock}

During propagation, part of the optical signal is lost by absorption, which transfers optical energy to the medium, and by scattering, which changes the propagation direction of the photons and can drive them away from the receiver. These processes are characterized by the absorption and scattering coefficients $a(\lambda)$ and $b(\lambda)$, and the extinction coefficient of the medium is $\alpha(\lambda)=a(\lambda)+b(\lambda)$ \cite{wozniak2007light}. The fraction of the signal retained after a path of length $\ell$ is given by the modified Beer-Lambert law \cite{elamassie2018performance}
\begin{equation}
\label{eq_transmittance}
A(\ell,\lambda)=\exp\!\left[-\alpha(\lambda)\left(\frac{d_1}{\theta_{\mathrm{rad}}}\right)^{T}\ell^{\,1-T}\right],\qquad A(0,\lambda)=1,
\end{equation}
where $d_1$ is the transmitter aperture diameter, $\theta_{\mathrm{rad}}$ the beam divergence angle and $T$ an empirical correction coefficient accounting for the finite beam divergence and the receiver aperture. Since this geometric correction is already included in Eq.~\eqref{eq_transmittance}, no additional geometric loss factor is applied. The overall transmission and detection efficiencies of the two arms are \cite{fung2006performance}
\begin{equation}
\label{eq_eta}
\eta_{A} = \eta_{\text{Alice}}\, A(x,\lambda),\qquad
\eta_{B} = \eta_{\text{Bob}}\, A(L - x,\lambda),
\end{equation}
where $\eta_{\text{Alice}}$ and $\eta_{\text{Bob}}$ are the overall efficiencies of Alice's and Bob's receivers. For $T\neq0$, the joint transmittance
\begin{equation}
\label{eq_joint_transmittance}
A(x,\lambda)A(L-x,\lambda)=\exp\!\left[-\alpha(\lambda)\left(\frac{d_1}{\theta_{\mathrm{rad}}}\right)^{T}S_T(x)\right],
\end{equation}
with $S_T(x)=x^{1-T}+(L-x)^{1-T}$, depends on $x$ even when $L$ is fixed, whereas for $T=0$ it reduces to $A(L,\lambda)$.
\end{revblock}
\vspace{0.2cm}

\item \textbf{Ambient irradiance}
\begin{revblock}

The ambient irradiance results from external sources such as sunlight or moonlight. Five illumination scenarios $s\in\{1,\dots,5\}$ are considered, with visible surface irradiances $E_{\mathrm{lit}}^{(s)}\in\{10^{-3},10,50,125,500\}$~W/m$^2$ \cite{paglierani2023primer} (see Section~\ref{Results}). These are irradiances integrated over the visible band and cannot be directly inserted in a narrowband photon-counting expression. To reproduce the background levels of the reference model of \cite{paglierani2023primer} within a consistent spectral and angular radiometric chain, equivalent irradiances $E_{\mathrm{vis,eq}}^{(s)}=F_{\mathrm{eq}}E_{\mathrm{lit}}^{(s)}$ are used, with $F_{\mathrm{eq}}\simeq1.0053\times10^{-5}$ obtained by equating the mean background photon numbers of both chains. The physical meaning of this factor is discussed in Section~\ref{sec_daylight}. Assuming a uniform spectral distribution over $[380,780]$~nm, the spectral irradiance at depth $z$ is \cite{mobley1994light}
\begin{equation}
\label{irradiance}
E_\lambda^{(s)}(z,\lambda)=\frac{E_{\mathrm{vis,eq}}^{(s)}}{400~\mathrm{nm}}\,e^{-K_\infty z},
\end{equation}
where $K_\infty$ is the diffuse attenuation coefficient of the ambient light field. The vertical depth $z$ must be distinguished from the propagation distance $\ell$, and $K_\infty$, which describes the vertical decay of ambient light, from $\alpha(\lambda)$, which describes the attenuation of the quantum signal along its path. Assuming a Lambertian downwelling light field, the spectral radiance is $L_{\lambda}^{(s)}=E_\lambda^{(s)}/\pi$.
\end{revblock}
\vspace{0.2cm}

\item \textbf{Photonic noise}
\begin{revblock}

Two independent noise sources are considered. Dark counts are false detections produced by the detectors in the absence of incident photons. Each of the $N_{\mathrm{det}}=4$ detectors of a receiver produces them at rate $I_{dc}$, giving a mean number $\overline{n}_{\mathrm{dc}}=N_{\mathrm{det}}I_{dc}\Delta t$ per bit period $\Delta t$. Background photons collected by a receiver of aperture area $S_{\mathrm{rx}}=\pi d_2^2/4$ during the acceptance window $\Delta t'$ have a mean number \cite{saleh2007fundamentals, rogers2006free}
\begin{equation}
\label{eq_nB}
n_B^{(s)}=\eta_{\mathrm{APD}}\,\frac{L_{\lambda}^{(s)}(z,\lambda)\,S_{\mathrm{rx}}\,\Omega_{\mathrm{eff}}\,\Delta\lambda\,\Delta t'\,\lambda}{h_p\,c_0},
\end{equation}
where $\Delta\lambda$ is the filter bandwidth, $h_p$ Planck's constant, $c_0$ the speed of light, $\eta_{\mathrm{APD}}$ the detector efficiency and
\begin{equation}
\label{eq_omega}
\Omega_{\mathrm{eff}}=\int_{\mathrm{FOV}}\cos\vartheta\,\mathrm{d}\Omega=\frac{\pi}{2}\sin^2\!\left(\frac{\delta_{\mathrm{rad}}}{2}\right)
\end{equation}
the projected angular acceptance for a full field of view $\delta_{\mathrm{rad}}$, the factor $1/2$ accounting for the overlap between the hemisphere in front of a horizontally pointing receiver and the downwelling hemisphere ($\Omega_{\mathrm{eff}}=\pi/2$~sr for $\delta=180^\circ$). Both processes being Poissonian, the probability of at least one noise click at a receiver during a window is
\begin{equation}
\label{equa_bruit}
y_0^{(s)}=1-\exp\!\left[-\overline{n}_{\mathrm{dc}}-n_B^{(s)}\right]\simeq \overline{n}_{\mathrm{dc}}+n_B^{(s)},
\end{equation}
where the approximation holds in the low-noise regime, whereas the exact expression is used in all numerical results. It is identical for Alice and Bob when both receivers are identical.
\end{revblock}
\end{enumerate}

\subsection{\chg{Coincidence probabilities}}
\begin{revblock}
A key bit is obtained only when both receivers click in the same window, i.e.\ for a coincidence. Denoting by $p_{\mathrm{pair}}$ the probability that a pair is emitted in a window (single-pair approximation), the click probability of receiver $X\in\{A,B\}$ is
\begin{equation}
\label{eq_click}
c_X=\eta_X+(1-\eta_X)\,y_0,
\end{equation}
where the second term is a noise click when the signal photon is not detected, which avoids double counting a signal click and a noise click at the same receiver. Detections being local and independent on the two arms, the coincidence probability is \cite{ma2007quantum}
\begin{equation}
\label{Proba coincidence}
\mathbb{P}_{\mathrm{coinc}}=p_{\mathrm{pair}}\,c_Ac_B+(1-p_{\mathrm{pair}})\,y_0^2.
\end{equation}
True coincidences, where both photons of the same pair are detected, occur with probability
\begin{equation}
\label{eq_ptrue}
\mathbb{P}_{\mathrm{true}}=p_{\mathrm{pair}}\,\eta_A\eta_B,
\end{equation}
and all other coincidences are accidental, with probability
\begin{align}
\label{eq_pfalse}
\mathbb{P}_{\mathrm{false}}&=\mathbb{P}_{\mathrm{coinc}}-\mathbb{P}_{\mathrm{true}}\\
&=p_{\mathrm{pair}}\big[\eta_A(1-\eta_B)y_0+\eta_B(1-\eta_A)y_0\notag\\
&\quad+(1-\eta_A)(1-\eta_B)y_0^2\big]+(1-p_{\mathrm{pair}})\,y_0^2.
\end{align}
The first two terms are signal-noise coincidences, in which the photon of one arm is detected while the click of the other arm is due to noise, whereas the last terms are noise-noise coincidences. As shown below, the signal-noise terms play a key role in the dependence of the performance on the source position.
\end{revblock}

\section{Quantum Channel Modeling with Kraus Operators}
\label{Section3}

The classical channel model introduced in Section~\ref{Classical channel} accounts for photon loss and background noise, but does not capture the quantum mechanical effects of the medium on the polarization state of the photons. A quantum description of the
propagation is therefore required to assess the impact of the underwater channel on the BBM92 protocol performance. In particular, the interaction between the photons and the scattering medium induces decoherence and irreversible polarization degradation, which progressively weakens the entanglement shared between Alice and Bob \cite{rizk:hal-05589560}.

\subsection{\chg{Open-system description and Kraus representation}}
\begin{revblock}
The source generates the pair in the state Eq.~\eqref{non-max-equation}, with density matrix
\begin{equation}
\label{matrice de densité in}
\small
\rho_{AB}^{\mathrm{in}}(\beta)
= \ket{\Phi_\beta}\bra{\Phi_\beta}
=
\begin{pmatrix}
\cos^{2}\beta & 0 & 0 & \cos\beta\,\sin\beta \\
0 & 0 & 0 & 0 \\
0 & 0 & 0 & 0 \\
\cos\beta\,\sin\beta & 0 & 0 & \sin^{2}\beta
\end{pmatrix}
\end{equation}
in the canonical basis $\{\ket{HH},\ket{HV},\ket{VH},\ket{VV}\}$ of $\mathcal{H}_{AB}=\mathcal{H}_A\otimes\mathcal{H}_B$. Without interaction with the medium, the photons would evolve under local unitaries $U_A\otimes U_B$, which preserve purity and entanglement. In an underwater channel, the photons interact with environmental degrees of freedom $E_A$ and $E_B$ that are neither controlled nor accessible to Alice and Bob \cite{breuer2007theory, helm2009quantum}. Assuming that these environments are initially uncorrelated with the pair and with each other, $\rho_{ABE_AE_B}^{\mathrm{in}}(\beta)=\rho_{AB}^{\mathrm{in}}(\beta)\otimes\rho_{E_A}\otimes\rho_{E_B}$, and that they interact locally with each photon through $U_{AE_A}\otimes U_{BE_B}$, the reduced state obtained by tracing out the environments is \cite{nielsen2010quantum}
\begin{align}
\label{eq_reduced_dynamics}
\rho_{AB}^{\mathrm{out}}(\beta)&=\mathrm{Tr}_{E_AE_B}\!\Big[U\big(\rho_{AB}^{\mathrm{in}}(\beta)\otimes\rho_{E_A}\otimes\rho_{E_B}\big)U^\dagger\Big]\\
&\equiv(\mathcal{E}_A\otimes\mathcal{E}_B)\big(\rho_{AB}^{\mathrm{in}}(\beta)\big),\qquad U=U_{AE_A}\otimes U_{BE_B}.
\end{align}
The local channels $\mathcal{E}_X$ are completely positive and trace-preserving (CPTP) maps \cite{wilde2013quantum, holevo2019quantum}, which admit the Kraus representation
\begin{equation}
\label{Kraus_oper_general}
\mathcal{E}_X(\rho_X)=\sum_i K_i^{(X)}\rho_X K_i^{(X)\dagger},\qquad \sum_i K_i^{(X)\dagger}K_i^{(X)}=\mathbb{I}_X,
\end{equation}
with at most four linearly independent operators for a qubit channel. The Kraus operators of the bipartite channel are $K_{ij}^{(AB)}=K_i^{(A)}\otimes K_j^{(B)}$, and they satisfy $\sum_{ij}K_{ij}^{(AB)\dagger}K_{ij}^{(AB)}=\mathbb{I}_A\otimes\mathbb{I}_B$.

Each photon is affected by two processes, namely a photon-loss channel, which determines whether the photon reaches the receiver, and a depolarizing channel, which degrades the polarization of the transmitted photons. As shown in Section~\ref{subsec_commutation}, these two channels commute in the proposed model, so that no ordering assumption is required.
\end{revblock}

\subsection{\chg{Photon-Loss Channel}}
\label{amplitude channel}
\begin{revblock}
In a quantum description, the loss of a photon is not a transformation between $\ket{H}$ and $\ket{V}$, since when a photon is absorbed or scattered out of the collected mode, no photon remains in this mode. This event is represented by the vacuum state $\ket{\mathrm{vac}}$, orthogonal to $\ket{H}$ and $\ket{V}$, so that the channel acts on the space spanned by $\{\ket{H},\ket{V},\ket{\mathrm{vac}}\}$. With $\tau_X=A(\ell_X,\lambda)$ the transmittance of arm $X$ given by Eq.~\eqref{eq_transmittance}, the photon-loss channel is described by the Kraus operators
\begin{equation}
\label{eq_kraus_loss}
\begin{aligned}
J_0^{(X)}&=\sqrt{\tau_X}\,\big(\ket{H}\bra{H}+\ket{V}\bra{V}\big)+\ket{\mathrm{vac}}\bra{\mathrm{vac}},\\
J_H^{(X)}&=\sqrt{1-\tau_X}\,\ket{\mathrm{vac}}\bra{H},\\
J_V^{(X)}&=\sqrt{1-\tau_X}\,\ket{\mathrm{vac}}\bra{V},
\end{aligned}
\end{equation}
which satisfy $\sum_k J_k^{(X)\dagger}J_k^{(X)}=\ket{H}\bra{H}+\ket{V}\bra{V}+\ket{\mathrm{vac}}\bra{\mathrm{vac}}=\mathbb{I}_X$\chg{, the term $\ket{\mathrm{vac}}\bra{\mathrm{vac}}$ in $J_0^{(X)}$ expressing that an absent photon remains absent}. For a normalized polarization state $\rho_X$, $\bra{H}\rho_X\ket{H}+\bra{V}\rho_X\ket{V}=1$, and
\begin{equation}
\label{eq_loss_action}
\mathcal{E}^{\mathrm{loss}}_X(\rho_X)=\tau_X\,\rho_X+(1-\tau_X)\ket{\mathrm{vac}}\bra{\mathrm{vac}}.
\end{equation}
For the pair, the joint action is $\mathcal{E}^{\mathrm{loss}}_A\otimes\mathcal{E}^{\mathrm{loss}}_B$. Projecting the output onto the subspace where both photons are present, with $\Pi_{AB}=\Pi_A\otimes\Pi_B$ and $\Pi_X=\ket{H}\bra{H}+\ket{V}\bra{V}$, gives
\begin{equation}
\Pi_{AB}\big(\mathcal{E}^{\mathrm{loss}}_A\otimes\mathcal{E}^{\mathrm{loss}}_B\big)(\rho_{AB}^{\mathrm{in}}(\beta))\,\Pi_{AB}=\tau_A\tau_B\,\rho_{AB}^{\mathrm{in}}(\beta),
\end{equation}
and, after normalization, the polarization state conditioned on the transmission of both photons is
\begin{equation}
\label{eq_rho_loss}
\rho_{AB}^{\mathrm{out,loss}}(\beta)=\rho_{AB}^{\mathrm{in}}(\beta).
\end{equation}
Hence, for polarization-independent losses, attenuation enters the model only through the efficiencies $\eta_A,\eta_B$ of Eq.~\eqref{eq_eta}, i.e.\ through the coincidence probabilities, and does not by itself create bit errors on the detected pairs.

This loss channel differs from the generalized amplitude-damping channel, whose Kraus operators map $\ket{V}$ onto $\ket{H}$ and which describes energy relaxation between two levels of the same system, with a thermal population $\xi$. Applied to polarization qubits, the latter would convert photon loss into polarization flips and thereby count the same loss twice, once in $\eta_{A,B}$ and once as an error. It is therefore not used here. Moreover, the mean thermal photon number at $\lambda=530$~nm and water temperature $T_{\mathrm{w}}\approx293$~K, $\bar n_{\mathrm{th}}=[\exp(h_pc_0/\lambda k_BT_{\mathrm{w}})-1]^{-1}\approx10^{-40}$, is negligible in any water body. Ambient light, including sunlight in shallow waters, is not a thermal input of the polarization mode but a source of uncorrelated detection events. It is fully accounted for through $n_B$ and $y_0$ in Eqs.~\eqref{eq_nB}--\eqref{equa_bruit} and studied in Section~\ref{sec_daylight}.
\end{revblock}

\subsection{Depolarizing Channel}
\label{depolarizing channel}
\chg{For the photons that reach the receivers, interactions with the medium may modify their polarization and reduce the correlations established by the source. This effect is represented by an isotropic depolarizing channel acting locally and independently on each photon}, in which the qubit state is preserved with probability $1-q$ and \chg{subjected to one of the three Pauli errors with probability $q/3$ each}, where $q \in [0,1]$ is
the depolarizing parameter. In contrast to a direct state flip that modifies the basis states, this channel degrades the quantum coherence of the superposition \chg{and can turn a pure state into a mixed state} \cite{de2020full}. The corresponding Kraus operators are \chg{\cite{nielsen2010quantum, sharma2015controlled}}
{\small
\begin{equation}
\begin{aligned}
L_0(q_X) &= \sqrt{1 - q_X}\, \mathbb{I}, \qquad
L_1(q_X) = \sqrt{\frac{q_X}{3}}\, \sigma_x,\\
L_2(q_X) &= \sqrt{\frac{q_X}{3}}\, \sigma_y,\qquad
L_3(q_X) = \sqrt{\frac{q_X}{3}}\, \sigma_z,
\end{aligned}
\end{equation}
}
where $\sigma_{x}, \sigma_{y}, \sigma_{z}$ are the
Pauli matrices\chg{, so that $\mathcal{E}^{\mathrm{dep}}_X(\rho_X)=(1-q_X)\rho_X+\frac{q_X}{3}\sum_{k}\sigma_k\rho_X\sigma_k$. The dependence of $q_X$ on the propagation length is parameterized as}
\begin{equation}
\label{eq_qX}
q_A = 1 - e^{-\gamma_{\mathrm{dep}} x},\qquad
q_B = 1 - e^{-\gamma_{\mathrm{dep}} (L-x)},
\end{equation}
\chg{where $\gamma_{\mathrm{dep}}$ is an effective depolarization coefficient given in Table~\ref{Table1}. For $q_X\ll1$, $q_X\simeq\gamma_{\mathrm{dep}}\ell_X$. The depolarization coefficient is estimated from the experiment of \cite{ji2017towards}, in which the polarization correlation of entangled pairs was measured before and after a propagation of $L_{\mathrm{exp}}=3$~m in seawater at $\lambda=810$~nm, with $V_{\mathrm{in}}=0.9528$ and $V_{\mathrm{out}}=0.9499$. When the depolarizing channel acts on one photon, the relation $\sigma_i\sigma_x\sigma_i=\sigma_x$ for $i=x$ and $-\sigma_x$ for $i\in\{y,z\}$ gives $V_{\mathrm{out}}=(1-\frac43q)V_{\mathrm{in}}$, so that
\begin{equation}
\label{eq_q_exp}
q=\frac34\left(1-\frac{V_{\mathrm{out}}}{V_{\mathrm{in}}}\right)\simeq2.28\times10^{-3}
\end{equation}
and, from Eq.~\eqref{eq_qX},
\begin{equation}
\label{eq_gamma_810}
\gamma_{\mathrm{dep}}(810\,\mathrm{nm})=-\frac{\ln(1-q)}{L_{\mathrm{exp}}}\simeq7.62\times10^{-4}~\mathrm{m}^{-1}.
\end{equation}
Since no measurement is available at $530$~nm, this value is extrapolated using the absorption spectrum of water. The ratio between the water absorption coefficients at $530$ and $810$~nm gives a reduction factor of about $10^{-2}$ \cite{wozniak2007light}. The extrapolated value is associated with turbid water, and $\gamma_{\mathrm{dep}}$ is assumed to vary between water types proportionally to their absorption coefficients $a_w=0.114$, $0.179$ and $0.366$~m$^{-1}$ for clear, coastal and turbid water \cite{mobley1994light}, so that
\begin{equation}
\label{eq_gamma_530}
\gamma_{\mathrm{dep}}^{(w)}(530\,\mathrm{nm})\simeq10^{-2}\,\gamma_{\mathrm{dep}}(810\,\mathrm{nm})\,\frac{a_w}{a_{\mathrm{turbid}}},
\end{equation}
where $a_{\mathrm{turbid}}=0.366$~m$^{-1}$ is the absorption coefficient of turbid water. This leads to $2.4\times10^{-6}$, $3.7\times10^{-6}$ and $7.6\times10^{-6}$~m$^{-1}$, as reported in Table~\ref{Table1}. Depolarization in a random medium also depends on the scattering phase function and on the particle size distribution, so that this scaling, based on absorption only, is approximate. These values must therefore be regarded as effective parameters giving an order of magnitude, and the robustness of the results to this approximation is assessed in Section~\ref{sec_gamma}.}

The bipartite channel has the 16 Kraus operators $L_{ij}=L_i(q_A)\otimes L_j(q_B)$, $i,j \in \{0,1,2,3\}$. \chg{Acting on the conditional state Eq.~\eqref{eq_rho_loss}, it gives}
{\small
\begin{revblock}
\begin{align}
\label{equa rho depo}
\rho_{AB}^{\text{out}}(\beta) &= \sum_{i,j=0}^{3} L_{ij} \,
\rho_{AB}^{\text{out,loss}}(\beta) \, L_{ij}^\dagger \\
&=\begin{pmatrix}
a_1 & 0 & 0 & k_1\\
0 & b_1 & 0 & 0\\
0 & 0 & c_1 & 0\\
k_1 & 0 & 0 & d_1
\end{pmatrix},
\end{align}
\end{revblock}}
\chg{where
\begin{equation}
\label{eq_coeffs}
\left\{
\begin{aligned}
a_1&=\tfrac14\big[1+r+(\kappa_A+\kappa_B)c_\beta\big],\\
b_1&=\tfrac14\big[1-r+(\kappa_A-\kappa_B)c_\beta\big],\\
c_1&=\tfrac14\big[1-r-(\kappa_A-\kappa_B)c_\beta\big],\\
d_1&=\tfrac14\big[1+r-(\kappa_A+\kappa_B)c_\beta\big],\\
k_1&=\tfrac12\,r\,s_\beta,
\end{aligned}
\right.
\end{equation}
with
\begin{equation}
\label{eq_params}
\kappa_X=1-\tfrac{4}{3}q_X,\quad r=\kappa_A\kappa_B,\quad c_\beta=\cos(2\beta),\quad s_\beta=\sin(2\beta).
\end{equation}

To derive Eq.~\eqref{equa rho depo}, we note that the Pauli operators form, together with the identity, a basis of the operators acting on a qubit. The input state Eq.~\eqref{matrice de densité in} can therefore be expanded as
\begin{align}
\label{eq_rho_pauli}
\rho_{AB}^{\mathrm{in}}(\beta)=\frac14\Big[&\mathbb{I}\otimes\mathbb{I}+c_\beta\,(\sigma_z\otimes\mathbb{I}+\mathbb{I}\otimes\sigma_z)+\sigma_z\otimes\sigma_z\notag\\
&+s_\beta\,(\sigma_x\otimes\sigma_x-\sigma_y\otimes\sigma_y)\Big],
\end{align}
where the coefficients follow from $\mathrm{Tr}[\rho_{AB}^{\mathrm{in}}(\beta)(\sigma_j\otimes\sigma_k)]$, which gives $\langle\sigma_z\otimes\mathbb{I}\rangle=\langle\mathbb{I}\otimes\sigma_z\rangle=\cos^2\beta-\sin^2\beta=c_\beta$, $\langle\sigma_z\otimes\sigma_z\rangle=1$, $\langle\sigma_x\otimes\sigma_x\rangle=-\langle\sigma_y\otimes\sigma_y\rangle=2\cos\beta\sin\beta=s_\beta$, all other terms being zero. For a single qubit, $\sigma_k\sigma_j\sigma_k=\sigma_j$ if $k=j$ and $-\sigma_j$ otherwise, so that $\sum_{k=1}^{3}\sigma_k\sigma_j\sigma_k=-\sigma_j$ and $\sum_{k=1}^{3}\sigma_k\,\mathbb{I}\,\sigma_k=3\,\mathbb{I}$. The local depolarizing channel therefore acts as
\begin{equation}
\mathcal{E}^{\mathrm{dep}}_X(\mathbb{I})=\mathbb{I},\qquad
\mathcal{E}^{\mathrm{dep}}_X(\sigma_j)=(1-q_X)\sigma_j-\frac{q_X}{3}\sigma_j=\kappa_X\,\sigma_j,
\end{equation}
i.e.\ it shrinks the Bloch vector of each photon by $\kappa_X$. By linearity of $\mathcal{E}^{\mathrm{dep}}_A\otimes\mathcal{E}^{\mathrm{dep}}_B$, each term $\sigma_j\otimes\mathbb{I}$ of Eq.~\eqref{eq_rho_pauli} is multiplied by $\kappa_A$, each term $\mathbb{I}\otimes\sigma_k$ by $\kappa_B$ and each term $\sigma_j\otimes\sigma_k$ by $r=\kappa_A\kappa_B$, which gives
\begin{align}
\rho_{AB}^{\mathrm{out}}(\beta)=\frac14\Big[&\mathbb{I}\otimes\mathbb{I}+c_\beta\,(\kappa_A\,\sigma_z\otimes\mathbb{I}+\kappa_B\,\mathbb{I}\otimes\sigma_z)+r\,\sigma_z\otimes\sigma_z\notag\\
&+r\,s_\beta\,(\sigma_x\otimes\sigma_x-\sigma_y\otimes\sigma_y)\Big].
\end{align}
Evaluating the matrix elements in the canonical basis, with $\sigma_z\ket{H}=\ket{H}$, $\sigma_z\ket{V}=-\ket{V}$ and
\begin{equation}
\sigma_x\otimes\sigma_x-\sigma_y\otimes\sigma_y=2\big(\ket{HH}\bra{VV}+\ket{VV}\bra{HH}\big),
\end{equation}
gives Eq.~\eqref{equa rho depo} with the coefficients of Eq.~\eqref{eq_coeffs}. For instance, $\bra{HH}\rho_{AB}^{\mathrm{out}}(\beta)\ket{HH}=\frac14[1+c_\beta(\kappa_A+\kappa_B)+r]=a_1$.

For a maximally entangled state, $\beta=\pi/4$, Eq.~\eqref{equa rho depo} reduces to the Werner-like state
\begin{equation}
\label{eq_rho_bell}
\rho_{AB}^{\mathrm{out}}\Big(\frac{\pi}{4}\Big)=\frac14\begin{pmatrix}1+r&0&0&2r\\0&1-r&0&0\\0&0&1-r&0\\2r&0&0&1+r\end{pmatrix},
\end{equation}
i.e.\ the mixture $r\ket{\Phi^+}\bra{\Phi^+}+(1-r)\,\mathbb{I}/4$.}

\subsection{\chg{Commutation of the loss and depolarizing channels}}
\label{subsec_commutation}
\begin{revblock}
To establish this property for any input state, including the entangled pair, both channels are written on the full space spanned by $\{\ket{H},\ket{V},\ket{\mathrm{vac}}\}$. With $\Pi_X=\ket{H}\bra{H}+\ket{V}\bra{V}$ and $P_X=\ket{\mathrm{vac}}\bra{\mathrm{vac}}$, any operator $\rho$ of arm $X$ is decomposed into the blocks $\rho_{pp}=\Pi_X\rho\Pi_X$, $\rho_{pv}=\Pi_X\rho P_X$, $\rho_{vp}=P_X\rho\Pi_X$ and $\rho_{vv}=P_X\rho P_X$, which describe the polarization part, the coherences between one photon and the vacuum, and the vacuum part. From Eq.~\eqref{eq_kraus_loss}, the loss channel gives
\begin{align}
\label{eq_loss_blocks}
\mathcal{E}^{\mathrm{loss}}_X(\rho)=\;&\tau_X\,\rho_{pp}+\sqrt{\tau_X}\,(\rho_{pv}+\rho_{vp})\notag\\
&+\rho_{vv}+(1-\tau_X)\,\mathrm{Tr}(\rho_{pp})\,P_X.
\end{align}
The depolarizing channel is extended to the vacuum by the Kraus operators $\tilde L_0=\sqrt{1-q_X}\,\Pi_X+P_X$ and $\tilde L_k=\sqrt{q_X/3}\,\sigma_k$, $k\in\{x,y,z\}$, where the Pauli operators act on the polarization subspace and vanish on the vacuum, so that $\sum_k\tilde L_k^\dagger\tilde L_k=\mathbb{I}_X$. Denoting by $\mathcal{D}_X(\sigma)=(1-q_X)\sigma+\frac{q_X}{3}\sum_k\sigma_k\sigma\sigma_k$ the qubit depolarizing map, which is linear and trace preserving, it gives
\begin{equation}
\label{eq_dep_blocks}
\mathcal{E}^{\mathrm{dep}}_X(\rho)=\mathcal{D}_X(\rho_{pp})+\sqrt{1-q_X}\,(\rho_{pv}+\rho_{vp})+\rho_{vv}.
\end{equation}
Applying Eq.~\eqref{eq_loss_blocks} to Eq.~\eqref{eq_dep_blocks}, or Eq.~\eqref{eq_dep_blocks} to Eq.~\eqref{eq_loss_blocks}, gives the same result in both cases,
\begin{align}
\label{eq_commutation}
\mathcal{E}^{\mathrm{loss}}_X\!\circ\mathcal{E}^{\mathrm{dep}}_X(\rho)&=\mathcal{E}^{\mathrm{dep}}_X\!\circ\mathcal{E}^{\mathrm{loss}}_X(\rho)\\
&=\tau_X\,\mathcal{D}_X(\rho_{pp})+\sqrt{\tau_X(1-q_X)}\,(\rho_{pv}+\rho_{vp})\notag\\
&\quad+\rho_{vv}+(1-\tau_X)\,\mathrm{Tr}(\rho_{pp})\,P_X,
\end{align}
since $\mathcal{D}_X(\tau_X\rho_{pp})=\tau_X\mathcal{D}_X(\rho_{pp})$ by linearity and $\mathrm{Tr}[\mathcal{D}_X(\rho_{pp})]=\mathrm{Tr}(\rho_{pp})$ by trace preservation. Eq.~\eqref{eq_commutation} holds for any operator $\rho$ and is therefore an equality between maps. For the pair, the property $(\mathcal{F}_A\otimes\mathcal{F}_B)\circ(\mathcal{G}_A\otimes\mathcal{G}_B)=(\mathcal{F}_A\circ\mathcal{G}_A)\otimes(\mathcal{F}_B\circ\mathcal{G}_B)$ then shows that the bipartite loss and depolarizing channels also commute, whatever the entanglement of $\rho_{AB}^{\mathrm{in}}(\beta)$.

The two channels therefore commute exactly, so that the correlations, QBER and SKR derived below do not depend on the order in which loss and depolarization are applied. This result is physically expected. Along the path, part of the photons is lost, by absorption or by scattering out of the field of view of the receiver, while the weakly deflected photons continue towards the receiver with a slightly modified polarization. These two phenomena occur simultaneously at each point of the path, so that no ordering is imposed by the medium. Each arm can therefore be described by a single channel, characterized by the transmittance $\tau_X$ and the depolarizing parameter $q_X$, which both depend on the arm length $\ell_X$. This property does not hold for the amplitude-damping description, which moves the Bloch vector towards $\ket{H}$, so that its composition with the depolarizing channel depends on the order.
\end{revblock}

\section{Performance Analysis of Non-Maximally Entangled State in BBM92}
\label{Section4}
\subsection{\chg{Quantum correlations}}
\label{sec_corr}
\begin{revblock}
The output density matrix of Section~\ref{Section3} contains all the information on the polarization state of the detected pairs. This section derives from it the quantum correlations, the QBER and the SKR. From Eq.~\eqref{equa rho depo}, the conditional correlations associated with the two measurement bases of BBM92 are
\begin{align}
C_X&=\langle\sigma_x\otimes\sigma_x\rangle=\mathrm{Tr}\big[\rho_{AB}^{\mathrm{out}}(\beta)(\sigma_x\otimes\sigma_x)\big]=r\sin(2\beta),\label{eq_CX}\\
C_Z&=\langle\sigma_z\otimes\sigma_z\rangle=\mathrm{Tr}\big[\rho_{AB}^{\mathrm{out}}(\beta)(\sigma_z\otimes\sigma_z)\big]=r.\label{eq_CZ}
\end{align}
Before depolarization, the photons of $\ket{\Phi_\beta}$ are perfectly correlated in the $H/V$ basis, whatever $\beta$, since the state only contains $\ket{HH}$ and $\ket{VV}$. The $D/A$ correlation, on the contrary, depends on the coherence between these two components and equals $\sin(2\beta)$, i.e.\ the concurrence Eq.~\eqref{eq_concurrence}. The depolarizing channel then reduces both correlations by the factor $r$. Substituting Eq.~\eqref{eq_qX} into Eq.~\eqref{eq_params}, the dependence of $C_X$ on the total distance and on the source position is
\begin{align}
\label{eq_CX_L}
C_X(L,x,\beta)=\frac{\sin(2\beta)}{9}&\left[4e^{-\gamma_{\mathrm{dep}}x}-1\right]\notag\\
\times&\left[4e^{-\gamma_{\mathrm{dep}}(L-x)}-1\right].
\end{align}
The degree of entanglement thus fixes the initial level of the correlation, while depolarization governs its decrease along the link.
\end{revblock}

\subsection{QBER analysis}
\label{QBER_analysis}
\begin{revblock}
The correlations derived above directly determine the bit errors of the true coincidences. For a correlation $C$ measured in a given basis, the probability that Alice and Bob obtain different results is $e=(1-C)/2$. The error probabilities of true coincidences in the $D/A$ and $H/V$ bases are therefore
\begin{equation}
e_X=\frac{1-r\sin(2\beta)}{2},\qquad e_Z=\frac{1-r}{2}.
\end{equation}

These expressions contain the error due to the non-maximality of the source. In the absence of depolarization ($r=1$), $e_Z=0$ and $e_X=\mathbb{P}_{\text{non-max}}$, where $\mathbb{P}_{\text{non-max}}$ is the probability of the discordant outcomes $\ket{DA}$ and $\ket{AD}$. Using $\braket{D}{H}=\braket{A}{H}=\braket{D}{V}=1/\sqrt2$ and $\braket{A}{V}=-1/\sqrt2$, the projection of Eq.~\eqref{non-max-equation} onto the diagonal basis gives
\begin{align}
\braket{DA}{\Phi_\beta}&=\cos\beta\,\braket{D}{H}\braket{A}{H}+\sin\beta\,\braket{D}{V}\braket{A}{V}\\
&=\frac{\cos\beta}{2}-\frac{\sin\beta}{2},
\end{align}
and, in the same way, $\braket{AD}{\Phi_\beta}=(\cos\beta-\sin\beta)/2$, so that
\begin{align}
\label{proba non max}
\mathbb{P}_{\text{non-max}}
&=|\braket{DA}{\Phi_\beta}|^2+|\braket{AD}{\Phi_\beta}|^2\\
&=\frac{(\cos\beta-\sin\beta)^2}{2}
=\frac{1 - \sin(2\beta)}{2}.
\end{align}
This error only occurs in the $D/A$ basis. The non-maximality of the source is therefore not an additional independent error mechanism, since it is contained in the correlation $C_X$ of Eq.~\eqref{eq_CX}, through the factor $\sin(2\beta)$, jointly with the channel contribution $r$. Treating it as an independent error, applied in both bases, would overestimate its contribution. Both bases being used with equal probability after sifting, the mean error probability of a true coincidence due to the source and the channel is
\begin{equation}
\label{eq_e_beta_dep}
e_{\beta,\mathrm{dep}}=\frac{e_X+e_Z}{2}=\frac{2-r\,[1+\sin(2\beta)]}{4},
\end{equation}
which reduces to $e_\beta=\mathbb{P}_{\text{non-max}}/2=[1-\sin(2\beta)]/4$ without depolarization and to $(1-r)/2$ for $\beta=\pi/4$.

The measurement system introduces an additional error with probability $e_{\mathrm{det}}$ (misalignment, imperfect polarization optics). Modeling $e_{\beta,\mathrm{dep}}$ and $e_{\mathrm{det}}$ as independent bit flips, a true coincidence yields an erroneous bit when exactly one of them occurs, which gives
\begin{align}
\label{proba canal+detector}
e_{\mathrm{sig}}&=e_{\beta,\mathrm{dep}}(1-e_{\mathrm{det}})+(1-e_{\beta,\mathrm{dep}})e_{\mathrm{det}}\\
&=e_{\mathrm{det}}+(1-2e_{\mathrm{det}})\,e_{\beta,\mathrm{dep}}.
\end{align}
Accidental coincidences contain at least one noise click, carry no correlation with the expected bit and produce an error with probability $1/2$. The QBER of BBM92 with a non-maximally entangled source over the considered quantum channel is finally
\begin{align}
\label{QBER_BBM92_non_max}
\text{QBER}_{\text{BBM92-non-max}}&=\frac{e_{\mathrm{sig}}\,\mathbb{P}_{\mathrm{true}}+\frac{1}{2}\mathbb{P}_{\mathrm{false}}}{\mathbb{P}_{\mathrm{coinc}}}\\
&=\frac12-\left(\frac12-e_{\mathrm{sig}}\right)\frac{\mathbb{P}_{\mathrm{true}}}{\mathbb{P}_{\mathrm{coinc}}}.
\end{align}
The second form shows that the QBER is governed by two quantities, namely the error of true coincidences $e_{\mathrm{sig}}$, which depends on the source and on depolarization, and the fraction of true coincidences $\mathbb{P}_{\mathrm{true}}/\mathbb{P}_{\mathrm{coinc}}$, which depends on losses and noise.

A minimum degree of entanglement is therefore required to extract a key. At short distance, where losses, noise and depolarization vanish, $\mathrm{QBER}\to e_{\mathrm{det}}+(1-2e_{\mathrm{det}})[1-\sin(2\beta)]/4$. A QBER below $11\,\%$ therefore requires
\begin{equation}
\label{eq_beta_min}
\mathcal{C}(\Phi_\beta)=\sin(2\beta)>1-\frac{4\,(0.11-e_{\mathrm{det}})}{1-2e_{\mathrm{det}}},
\end{equation}
i.e.\ $\mathcal{C}>0.670$ and $\beta>21.04^\circ$ for $e_{\mathrm{det}}=3.3\,\%$, whatever the water type. Below this degree of entanglement, no secret key can be extracted at any distance.
\end{revblock}

\subsection{SKR analysis}
\label{SKR_analysis}
\begin{revblock}
While the QBER indicates whether a secret key can be extracted, the SKR quantifies the number of secure bits extracted per generated pair. Extracting a secret key from the raw key requires error correction and privacy amplification. In the asymptotic regime, the secret fraction is given by the Devetak-Winter bound \cite{devetak2005distillation}, $1-f_{\mathrm{EC}}h_2(\mathrm{QBER})-\chi_E$, where $h_2(p)=-p\log_2p-(1-p)\log_2(1-p)$ is the binary entropy, $f_{\mathrm{EC}}h_2(\mathrm{QBER})$ the error-correction cost and $\chi_E=h_2(\mathrm{QBER})$ the upper bound on the information of an eavesdropper, removed by privacy amplification \cite{shor2000simple}. Following \cite{paglierani2023primer}, the reconciliation efficiency is related to the code rate $R_c$ by $f_{\mathrm{EC}}=(1-R_c)/h_2(0.11)$, with $R_c=0.5$ for the considered Low-Density Parity-Check (LDPC) code \cite{5205475}, which gives $f_{\mathrm{EC}}\simeq1$. Since only half of the coincidences survive sifting, the SKR is \cite{waks2002security, muskan2023analysing}
\begin{align}
\label{eq_SKR}
\text{SKR}_{\text{BBM92-non-max}}=\frac{\mathbb{P}_{\mathrm{coinc}}}{2}\,
\max\!\Big[0,\;&1-\left(1+f_{\mathrm{EC}}\right)\notag\\
&\times h_2\big(\text{QBER}_{\text{BBM92-non-max}}\big)\Big].
\end{align}
With $f_{\mathrm{EC}}\simeq1$, the secret fraction vanishes for a QBER of about $11\,\%$, in agreement with the threshold of Section~\ref{Section 2}. Finite-size effects are not considered.
\end{revblock}

\section{Results and Discussion}
\label{Results}
We analyse the \chg{correlations,} QBER and SKR of the BBM92 protocol with
non-maximally entangled states over a non-turbulent underwater channel, across three water types, namely clear, coastal and turbid water, following Mobley's classification \cite{mobley1994light}, and five atmospheric conditions \cite{raouf2022performance, paglierani2023primer}, described as follows.
\begin{itemize}
    \item Scenario~1 corresponds to a clear atmosphere with a full moon near the zenith \chg{($E_{\mathrm{lit}}^{(1)}=10^{-3}$~W/m$^2$)}.
    \item Scenario~2 represents a heavy overcast with the sun near the horizon \chg{($E_{\mathrm{lit}}^{(2)}=10$~W/m$^2$)}.
    \item Scenario~3 models a hazy atmosphere with the sun near the horizon \chg{($E_{\mathrm{lit}}^{(3)}=50$~W/m$^2$)}.
    \item Scenario~4 refers to a heavy overcast with the sun at the zenith \chg{($E_{\mathrm{lit}}^{(4)}=125$~W/m$^2$)}.
    \item Scenario~5 describes a clear atmosphere with the sun at the zenith \chg{($E_{\mathrm{lit}}^{(5)}=500$~W/m$^2$)}.
\end{itemize}

\begin{table}[!t]
\centering
\caption{System and channel parameters}
\label{Table1}
\begin{tabular}{@{}c|l|c@{}}
\toprule
Parameter & Definition & Value \\
\midrule
$\lambda$        & Wavelength \cite{paglierani2023primer}               & 530 nm \\
\chg{$\theta$}   & \chg{Beam divergence angle}                          & \chg{$6^\circ$} \\
\chg{$\delta$}   & \chg{Receiver full field of view} \cite{paglierani2023primer} & \chg{$180^\circ$} \\
$\Delta \lambda$ & Filter bandwidth \cite{ata2025impact}                 & 0.2 nm\\
$\Delta t $      & \chg{Bit period} \cite{ata2025impact}                 & 40 ns\\
$\Delta t'$      & Receiver gate time \cite{ata2025impact}               & 200 ps\\
\chg{$d_1$, $d_2$} & \chg{Transmitter / receiver apertures}              & \chg{10, 30 cm} \\
\chg{$\eta_{\mathrm{APD}}$} & \chg{Detector efficiency} \cite{paglierani2023primer} & \chg{0.5} \\
$\eta_{\text{Alice}}$, $\eta_{\text{Bob}}$ & Receiver efficiencies   & 0.5\\
\chg{$p_{\mathrm{pair}}$} & \chg{Pair probability per window}           & \chg{1} \\
$I_{dc}$         & Dark count rate \cite{paglierani2023primer}           & 60 Hz\\
$z$              & Depth \cite{ata2025impact}                            & 80 m \\
$K_\infty$       & Diffuse attenuation coeff. \cite{paglierani2023primer} & 0.08 m$^{-1}$\\
$e_\text{det}$   & Detection error rate \cite{paglierani2023primer}      & 3.3\% \\
\chg{$R_c$}      & \chg{LDPC code rate} \cite{5205475}                   & \chg{0.5} \\
\midrule
$\alpha$         & Extinction coefficient \cite{mobley1994light}         &  \\[-2pt]
                 & \qquad Clear water                                    & 0.151 m$^{-1}$ \\
                 & \qquad Coastal water                                  & \chg{0.398} m$^{-1}$ \\
                 & \qquad Turbid water                                   & \chg{2.190} m$^{-1}$ \\
\midrule
$T$              & Correction coefficient \cite{paglierani2023primer}    &  \\[-2pt]
                 & \qquad \chg{$d_1=d_2=10$ cm}                          & 0.16 \\
                 & \qquad \chg{$d_1=d_2=30$ cm}                          & \chg{0.26} \\
\midrule
$\gamma_\text{dep}$ & Depolarization coefficient                         &  \\[-2pt]
                 & \qquad Clear water                                    & $2.4\times 10^{-6}$ m$^{-1}$ \\
                 & \qquad Coastal water                                  & $3.7\times 10^{-6}$ m$^{-1}$ \\
                 & \qquad Turbid water                                   & \chg{$7.6\times 10^{-6}$} m$^{-1}$ \\
\bottomrule
\end{tabular}
\end{table}

\begin{revblock}
The parameters are summarized in Table~\ref{Table1} \cite{paglierani2023primer, ata2025impact}. Unless otherwise stated, two reference configurations are considered, namely $(i)$ a maximally entangled source ($\beta=45^\circ$) placed at the centre of the link ($x=0.5L$) with $d_1=d_2=30$~cm, and $(ii)$ a non-maximally entangled source ($\beta=30^\circ$, $\mathcal{C}\simeq0.866$) placed at $x=0.3L$ with $d_1=d_2=10$~cm, which combines a lower entanglement, smaller apertures and an asymmetric deployment. The value $p_{\mathrm{pair}}=1$ is an idealized reference describing the elementary coincidence dynamics and should not be interpreted as the generation probability of a real spontaneous parametric down-conversion source. The threshold distance $L_{\mathrm{th}}$ is defined as the largest distance for which the QBER remains below $11\,\%$. It is used as a comparative indicator and does not include finite-size effects.

The analytical expressions are compared with Monte Carlo simulations performed at the level of the emission windows \cite{chalupnik2025realistic}. For each distance, $300$ independent realizations are generated. In each of them, $N_{\mathrm{win}}=\max\!\left(10^4,\lceil10^3/\mathbb{P}_{\mathrm{coinc}}\rceil\right)$ windows are distributed by a multinomial draw among five exclusive categories, namely correct and erroneous true coincidences, with probabilities $\mathbb{P}_{\mathrm{true}}(1-e_{\mathrm{sig}})$ and $\mathbb{P}_{\mathrm{true}}e_{\mathrm{sig}}$, correct and erroneous accidental coincidences, each with probability $\mathbb{P}_{\mathrm{false}}/2$, and no coincidence. The adaptive number of windows keeps about $10^3$ coincidences per realization when $\mathbb{P}_{\mathrm{coinc}}$ becomes small. The simulated QBER is the ratio of erroneous to total coincidences, and the simulated SKR is obtained from the simulated coincidence rate and QBER through Eq.~\eqref{eq_SKR}. For the correlations, $10^7$ pairs are measured per realization, each pair giving concordant outcomes with probability $(1+C_X)/2$. Markers and error bars in the figures give the mean and standard deviation over the realizations.
\end{revblock}

\subsection{\chg{Quantum correlations}}

\begin{figure}[!htbp]
    \centering
    \includegraphics[width=0.88\linewidth]{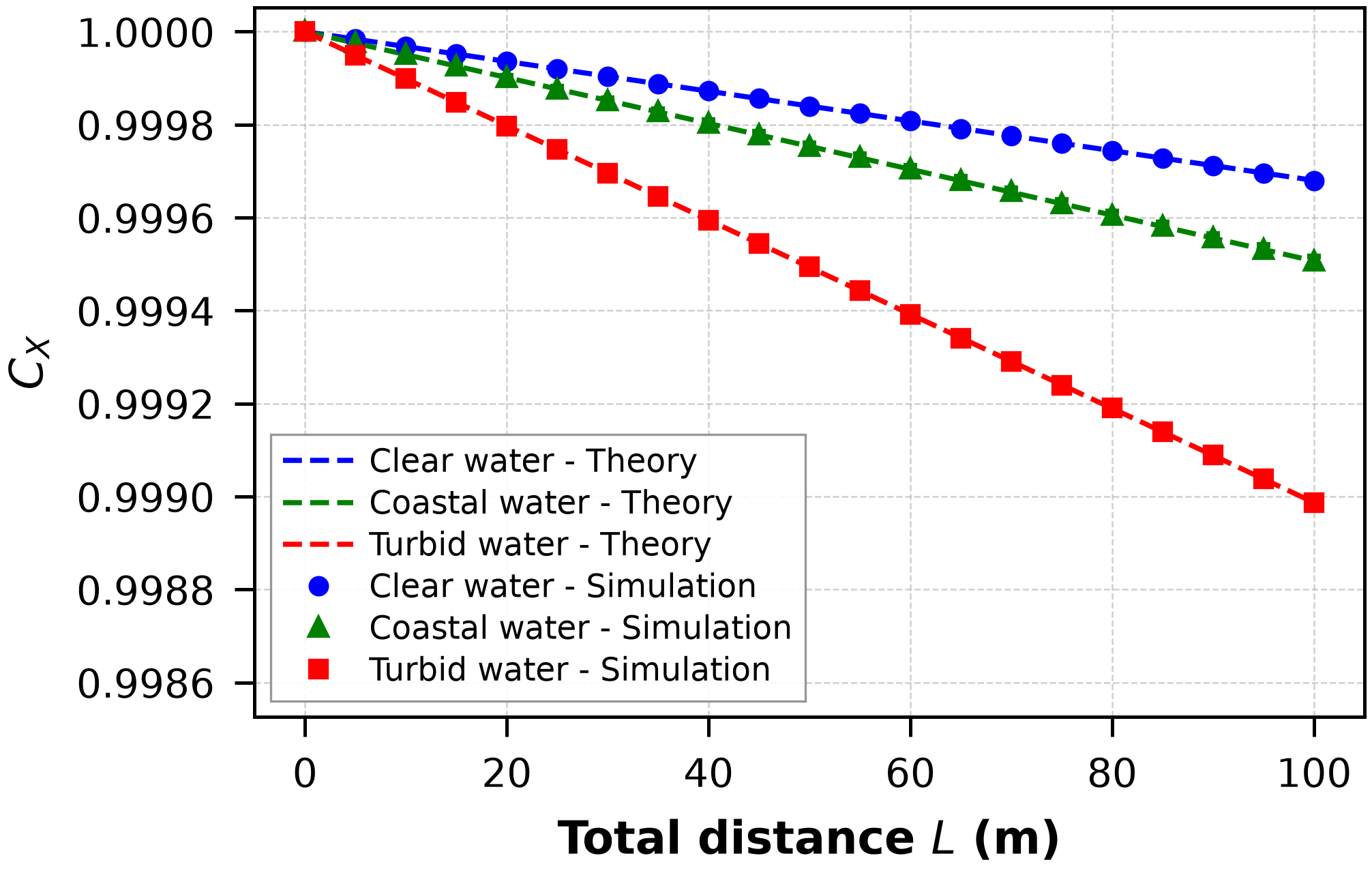}\\[-2pt]
    {\small (a)}\\[5pt]
    \includegraphics[width=0.88\linewidth]{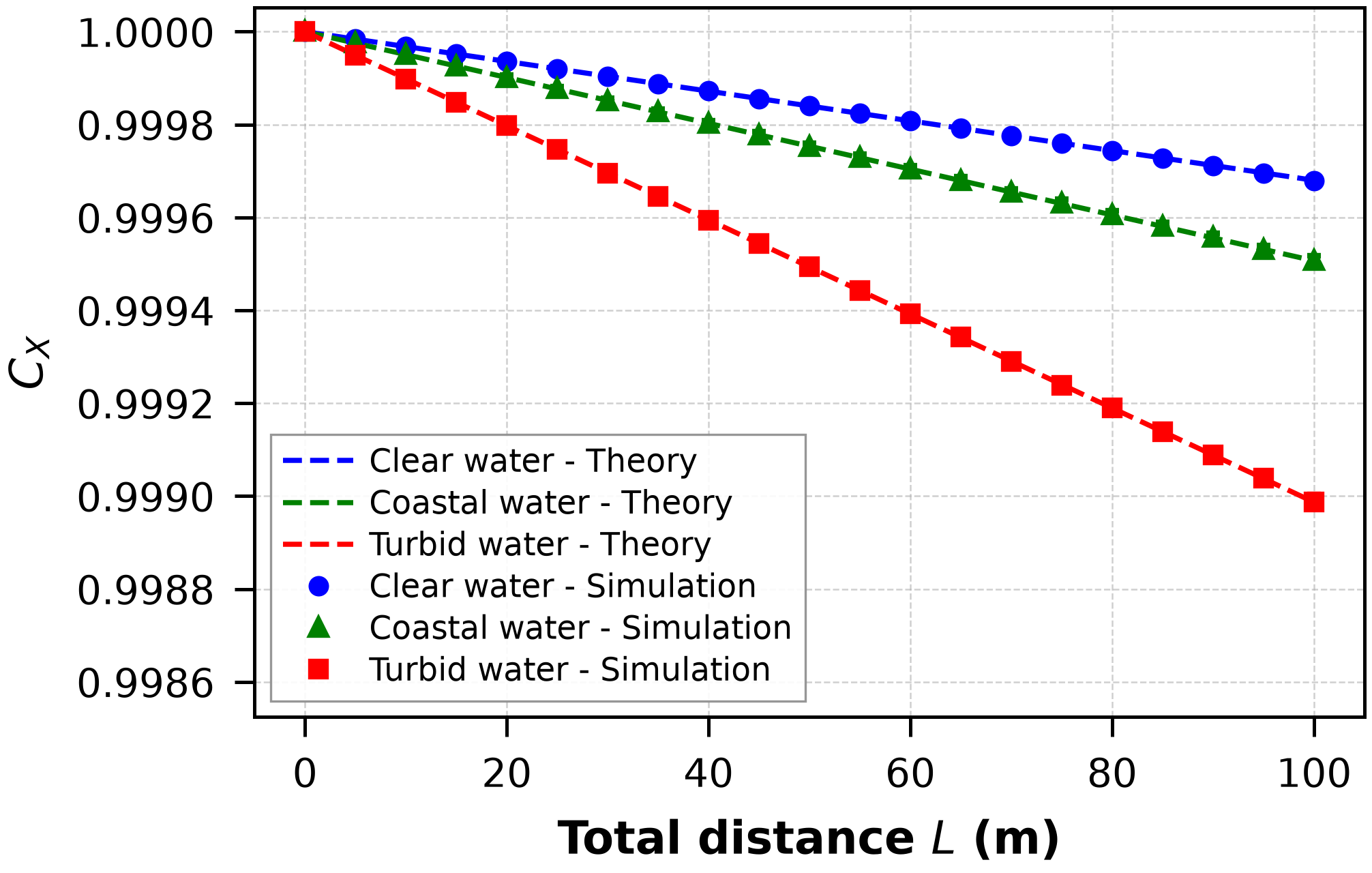}\\[-2pt]
    {\small (b)}\\[5pt]
    \includegraphics[width=0.88\linewidth]{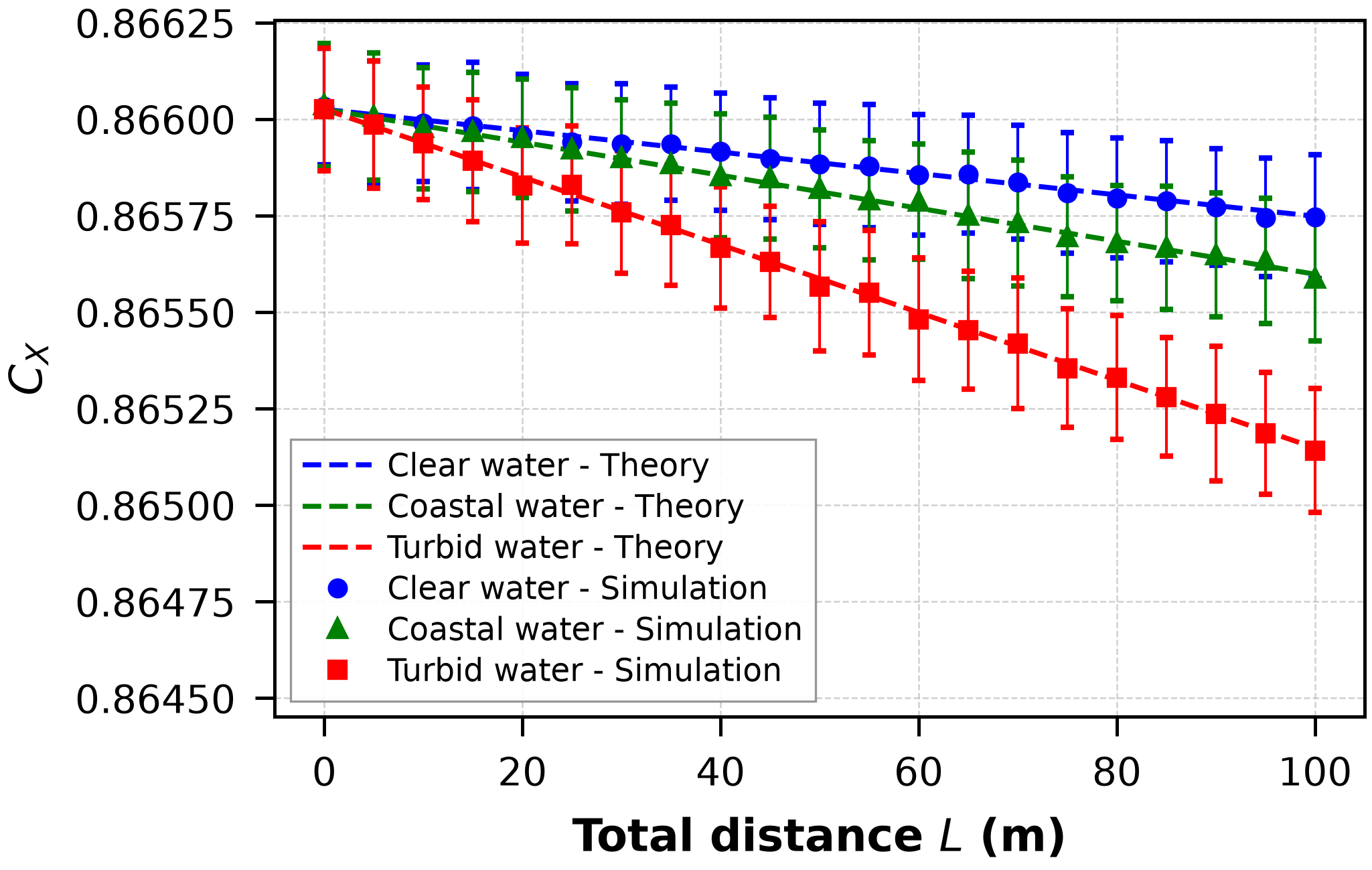}\\[-2pt]
    {\small (c)}\\[5pt]
    \includegraphics[width=0.88\linewidth]{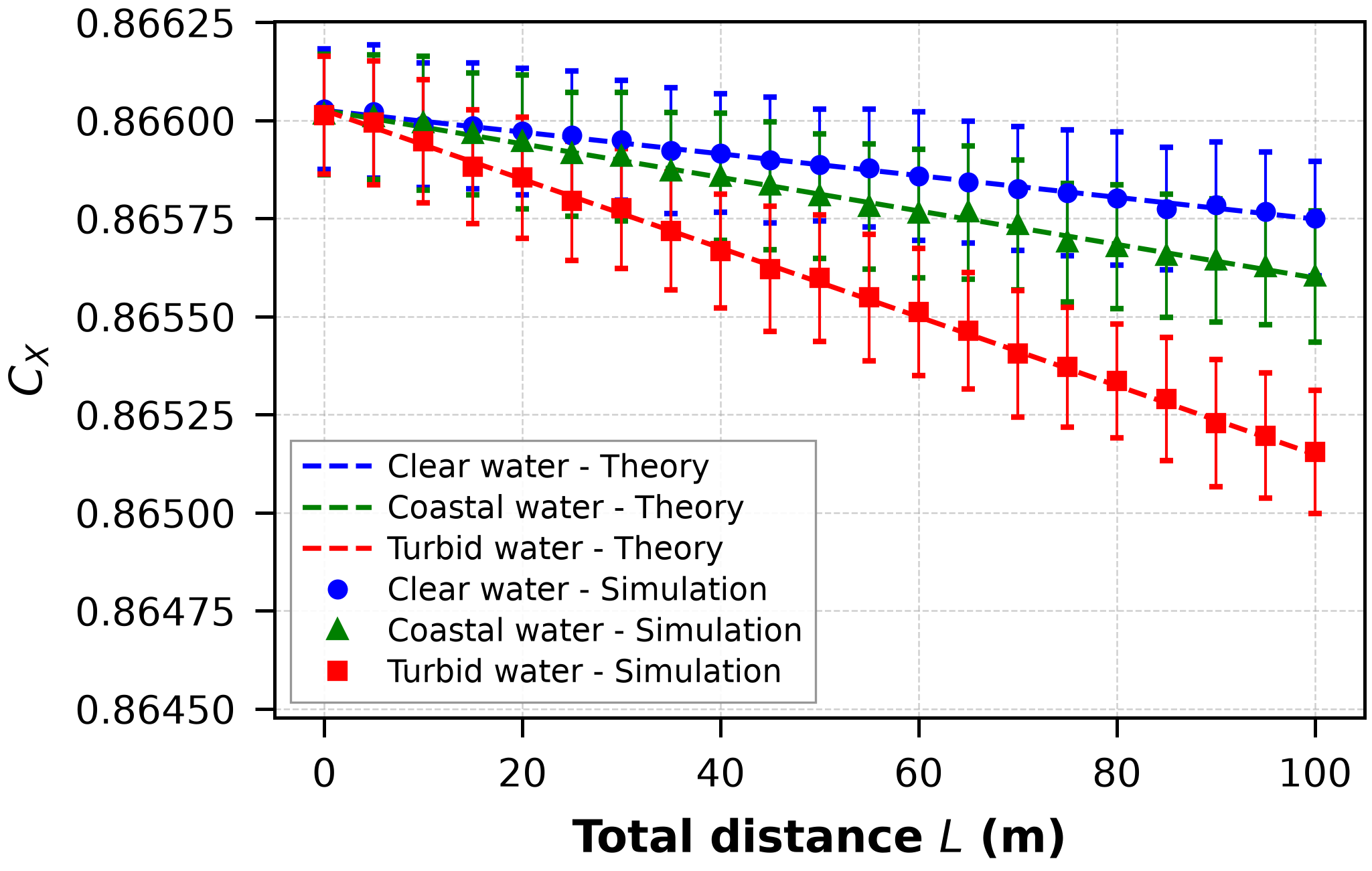}\\[-2pt]
    {\small (d)}
    \caption{\small \chg{Conditional correlation $C_X=\langle\sigma_x\otimes\sigma_x\rangle$ versus the total distance $L$ in clear, coastal and turbid water, for (a) $\beta=45^\circ$, $x=0.5L$, $d_1=d_2=30$~cm, (b) $\beta=45^\circ$, $x=0.2L$, $d_1=d_2=30$~cm, (c) $\beta=30^\circ$, $x=0.5L$, $d_1=d_2=10$~cm and (d) $\beta=30^\circ$, $x=0.3L$, $d_1=d_2=10$~cm. Dashed lines represent the analytical expression of Eq.~\eqref{eq_CX_L} and markers the Monte Carlo simulations. Since loss does not modify the conditional polarization state, $C_X$ does not depend on the aperture diameters.}}
\label{fig_corr}
\end{figure}

\begin{revblock}
The analysis starts with the quantum correlations, which isolate the effect of the source and of depolarization, since losses do not modify the polarization state of the detected pairs. Fig.~\ref{fig_corr} shows the evolution of the diagonal-basis correlation $C_X$ with distance, for $\beta=45^\circ$ with $d_1=d_2=30$~cm and for $\beta=30^\circ$ with $d_1=d_2=10$~cm, i.e.\ the two reference configurations used for the QBER and SKR. In Figs.~\ref{fig_corr}(a) and \ref{fig_corr}(b), which correspond to the maximally entangled state, $C_X$ starts at unity and its decrease directly reflects the depolarization accumulated along the link. This decrease is smallest in clear water and increases in coastal and turbid water, following the increase of $\gamma_{\mathrm{dep}}$. After $100$~m, $C_X=0.99968$, $0.99951$ and $0.99899$, respectively. The results for $x=0.5L$ and $x=0.2L$ are almost identical, because for the small values of $\gamma_{\mathrm{dep}}$, $r\simeq1-\frac43\gamma_{\mathrm{dep}}L$ depends at first order only on the total length $L$, not on its distribution between the two arms. In Figs.~\ref{fig_corr}(c) and \ref{fig_corr}(d), which correspond to $\beta=30^\circ$, the correlation starts at $\sin(60^\circ)\simeq0.866$ instead of unity. This offset results from the non-maximality of the emitted state and not from the channel. Over the whole range, the correlation for $\beta=45^\circ$ remains larger than for $\beta=30^\circ$, and the same ordering of the water types is observed. These results show that the correlation is governed by two mechanisms, since the degree of entanglement at emission sets its initial level, whereas the depolarizing channel determines its degradation during propagation. They also show that, for the considered parameters, depolarization is a weak effect over distances of the order of $100$~m.
\end{revblock}

\subsection{\chg{QBER for a maximally entangled source}}

\begin{figure}[!htbp]
    \centering
    \includegraphics[width=1.0\linewidth]{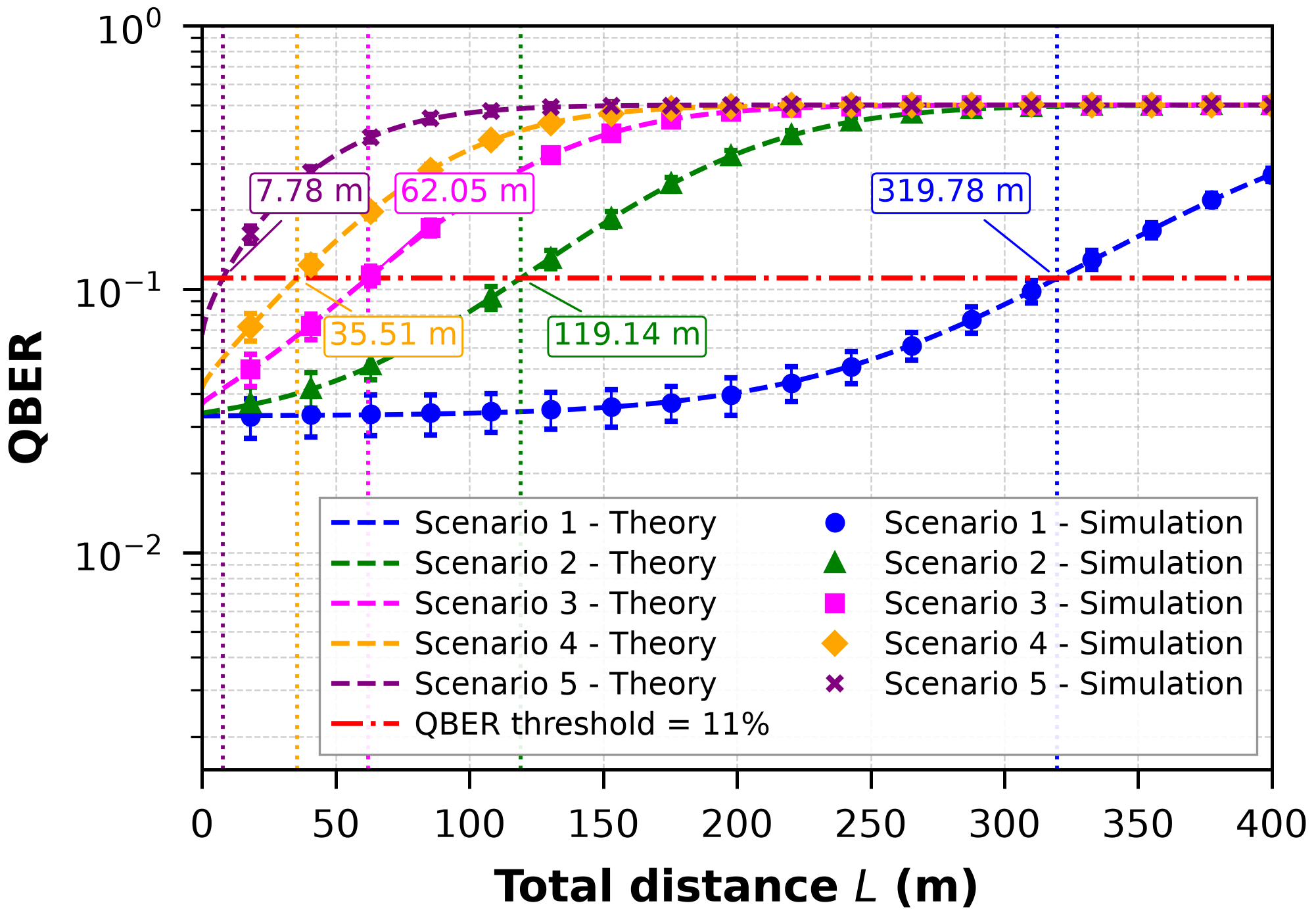}\\[-2pt]
    {\small (a)}\\[5pt]
    \includegraphics[width=1.0\linewidth]{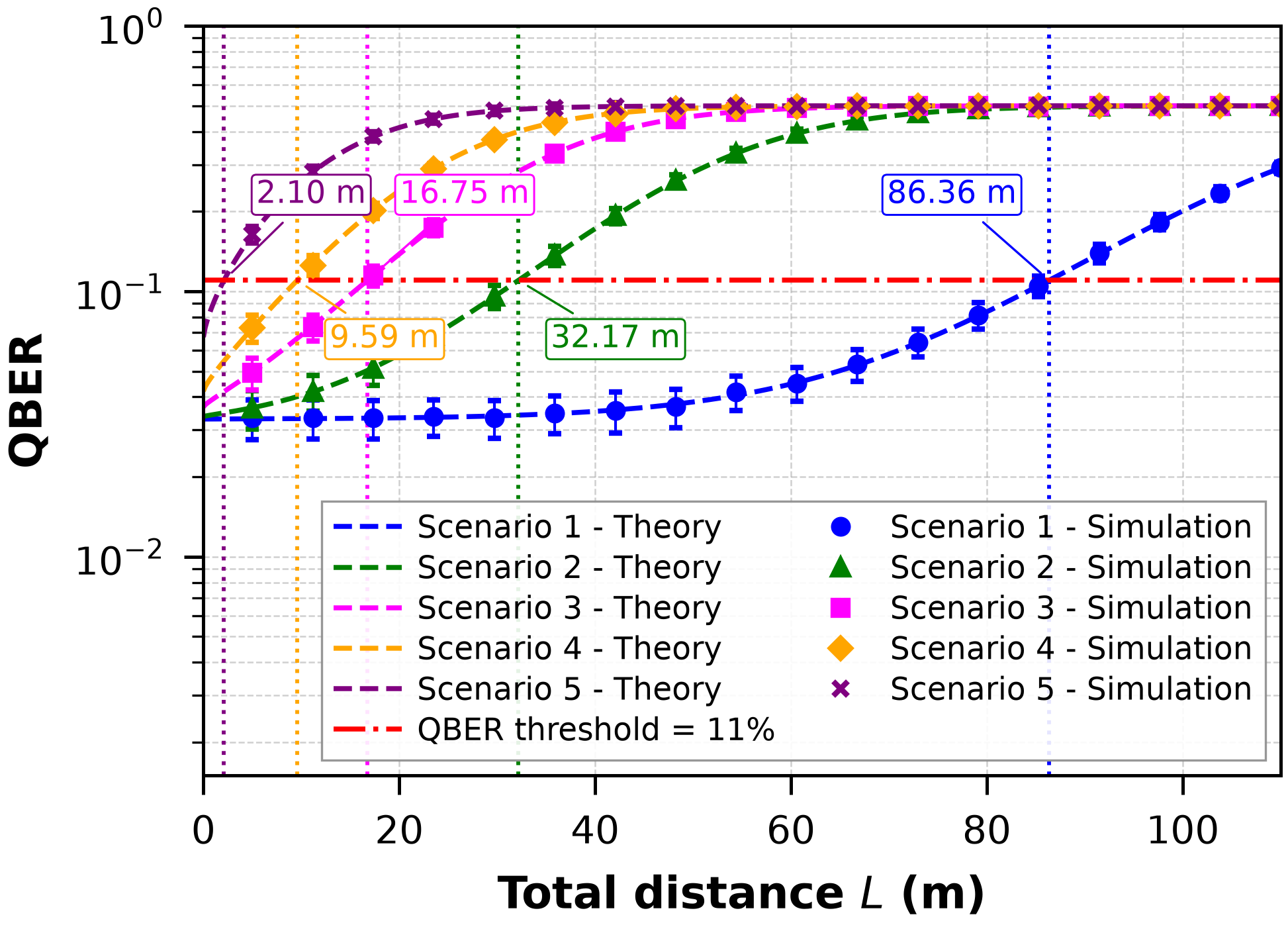}\\[-2pt]
    {\small (b)}\\[5pt]
    \includegraphics[width=1.0\linewidth]{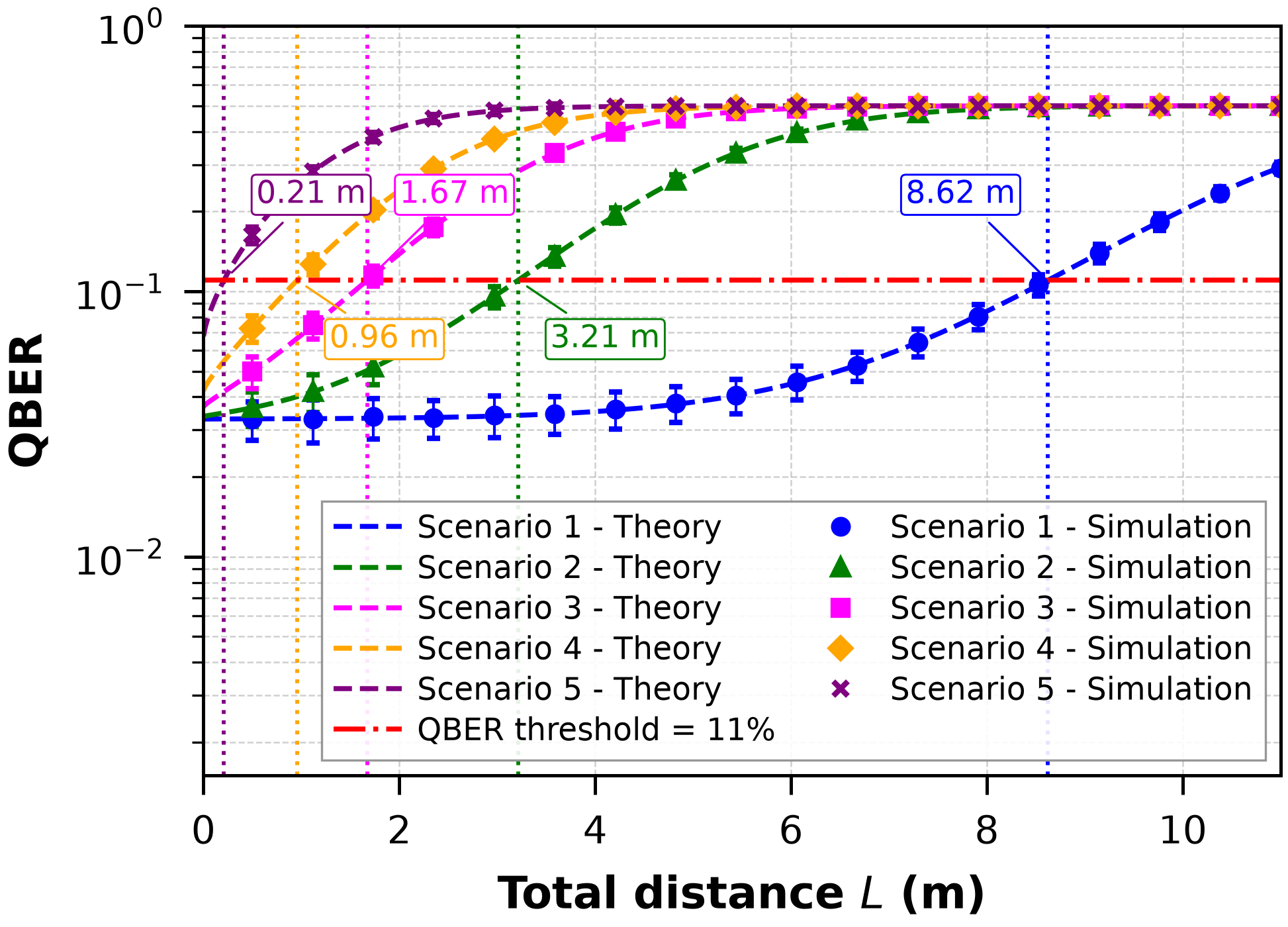}\\[-2pt]
    {\small (c)}
    \caption{\small \chg{QBER of BBM92 versus the total distance $L$ for $\beta=45^\circ$, $x=0.5L$ and $d_1=d_2=30$~cm, under the five irradiance scenarios, in (a) clear water, (b) coastal water and (c) turbid water. Dashed lines represent the analytical model, markers the Monte Carlo simulations, and the labels give the threshold distances.}}
\label{figure 1}
\end{figure}

\begin{revblock}
The correlations only describe the pairs that are actually detected. The QBER also accounts for losses and background noise through the fraction of true coincidences, and it is therefore the quantity that determines the secure range. Fig.~\ref{figure 1} shows the QBER of BBM92 as a function of distance for $\beta=45^\circ$, $x=0.5L$ and $d_1=d_2=30$~cm. The analytical results agree with the Monte Carlo simulations in all configurations. At short distances, the QBER is close to $e_{\mathrm{det}}=3.3\,\%$. It then increases with distance, as the true coincidences decrease and accidental coincidences become dominant, and it saturates at $50\,\%$. For a given scenario, this increase occurs at shorter distances from clear to coastal and turbid water.

In Fig.~\ref{figure 1}(a), which corresponds to clear water, the $11\,\%$ threshold is reached at $319.78$, $119.14$, $62.05$, $35.51$ and $7.78$~m for Scenarios~1 to~5, respectively. These distances decrease to $86.36$, $32.17$, $16.75$, $9.59$ and $2.10$~m in coastal water, as shown in Fig.~\ref{figure 1}(b), and to $8.62$, $3.21$, $1.67$, $0.96$ and $0.21$~m in turbid water, as shown in Fig.~\ref{figure 1}(c). Ambient irradiance thus strongly reduces the range, and this reduction becomes more pronounced as the attenuation of the medium increases. Since the transmittance Eq.~\eqref{eq_transmittance} depends on $\alpha\,\ell^{1-T}$, the threshold distances scale as $\alpha^{-1/(1-T)}$ for a given scenario. The ratio between clear and turbid water, $(2.190/0.151)^{1/0.74}\simeq37$, is indeed recovered, since $319.78/8.62\simeq37.1$.

Setting $\gamma_{\mathrm{dep}}=0$ in the same configuration, which corresponds to a classical link-budget model, gives $320.09$~m instead of $319.78$~m in clear water and $86.39$~m instead of $86.36$~m in coastal water under Scenario~1, and $35.53$~m instead of $35.51$~m in clear water under Scenario~4. For the considered parameters, depolarization is therefore a weak effect compared with attenuation and background noise, and the quantum description allows these contributions to be separated and the QBER variations to be related to the evolution of the two-photon state.
\end{revblock}

\subsection{\chg{QBER for a non-maximally entangled source}}

\begin{figure}[!htbp]
    \centering
    \includegraphics[width=1.0\linewidth]{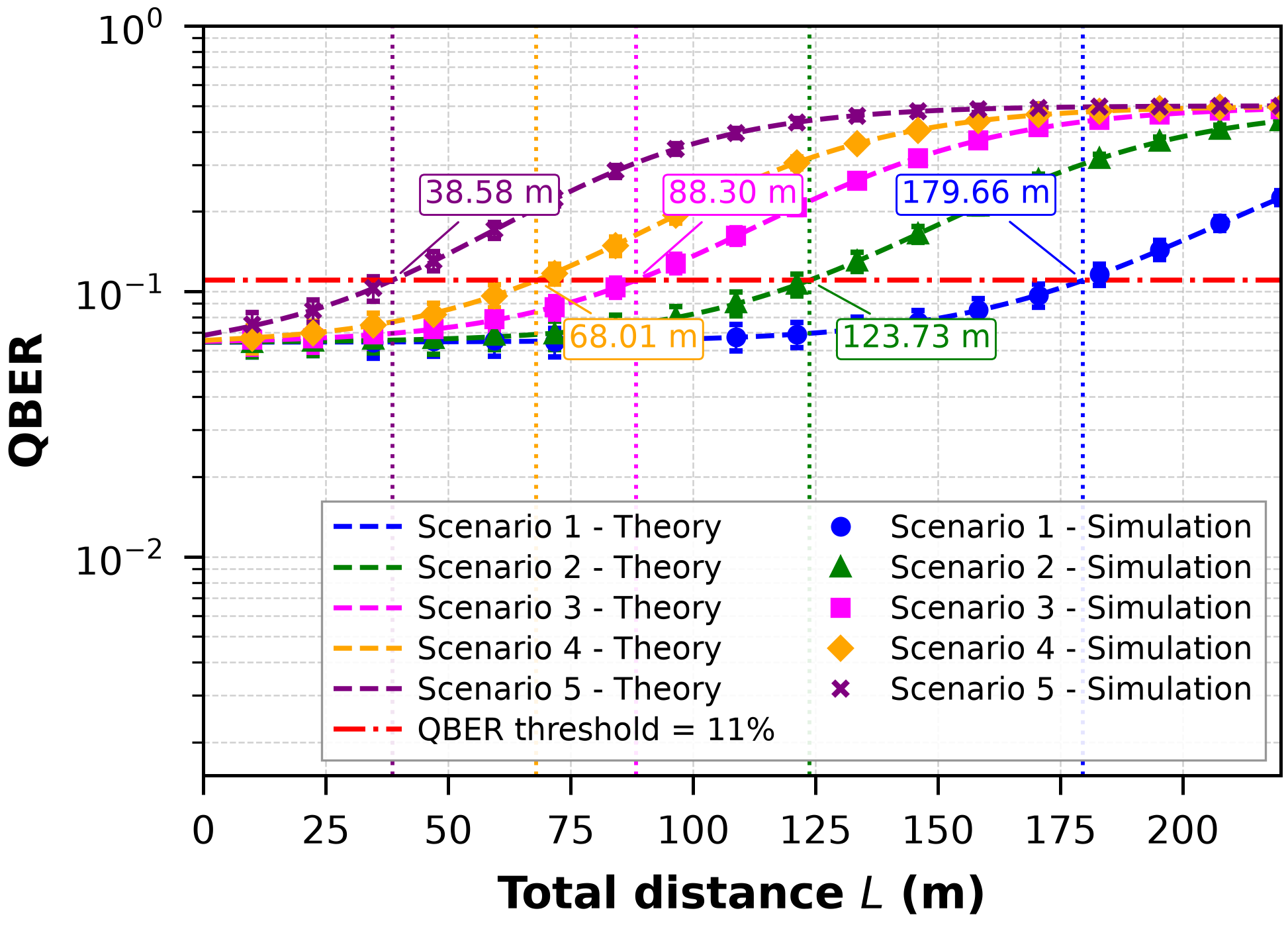}\\[-2pt]
    {\small (a)}\\[5pt]
    \includegraphics[width=1.0\linewidth]{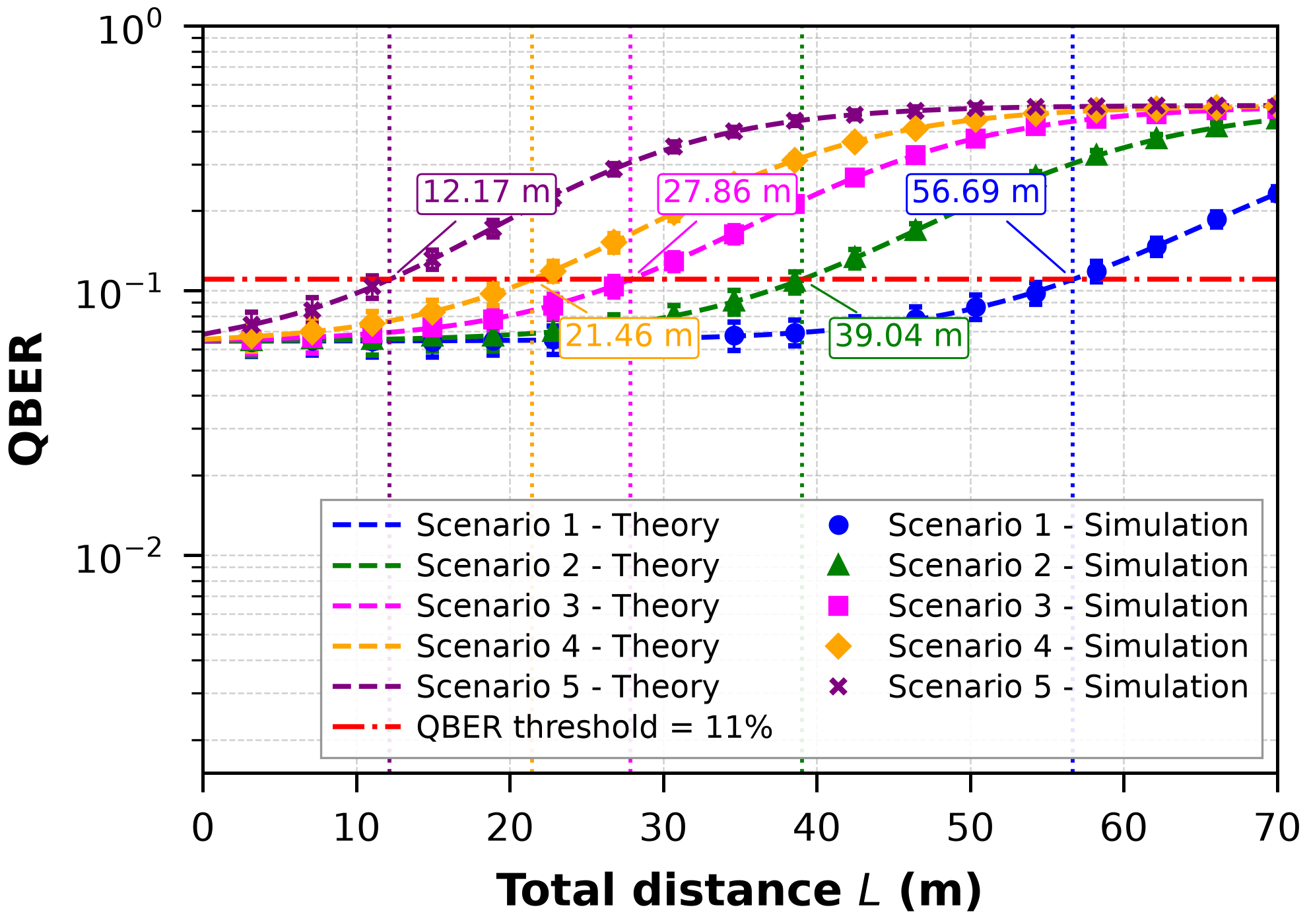}\\[-2pt]
    {\small (b)}\\[5pt]
    \includegraphics[width=1.0\linewidth]{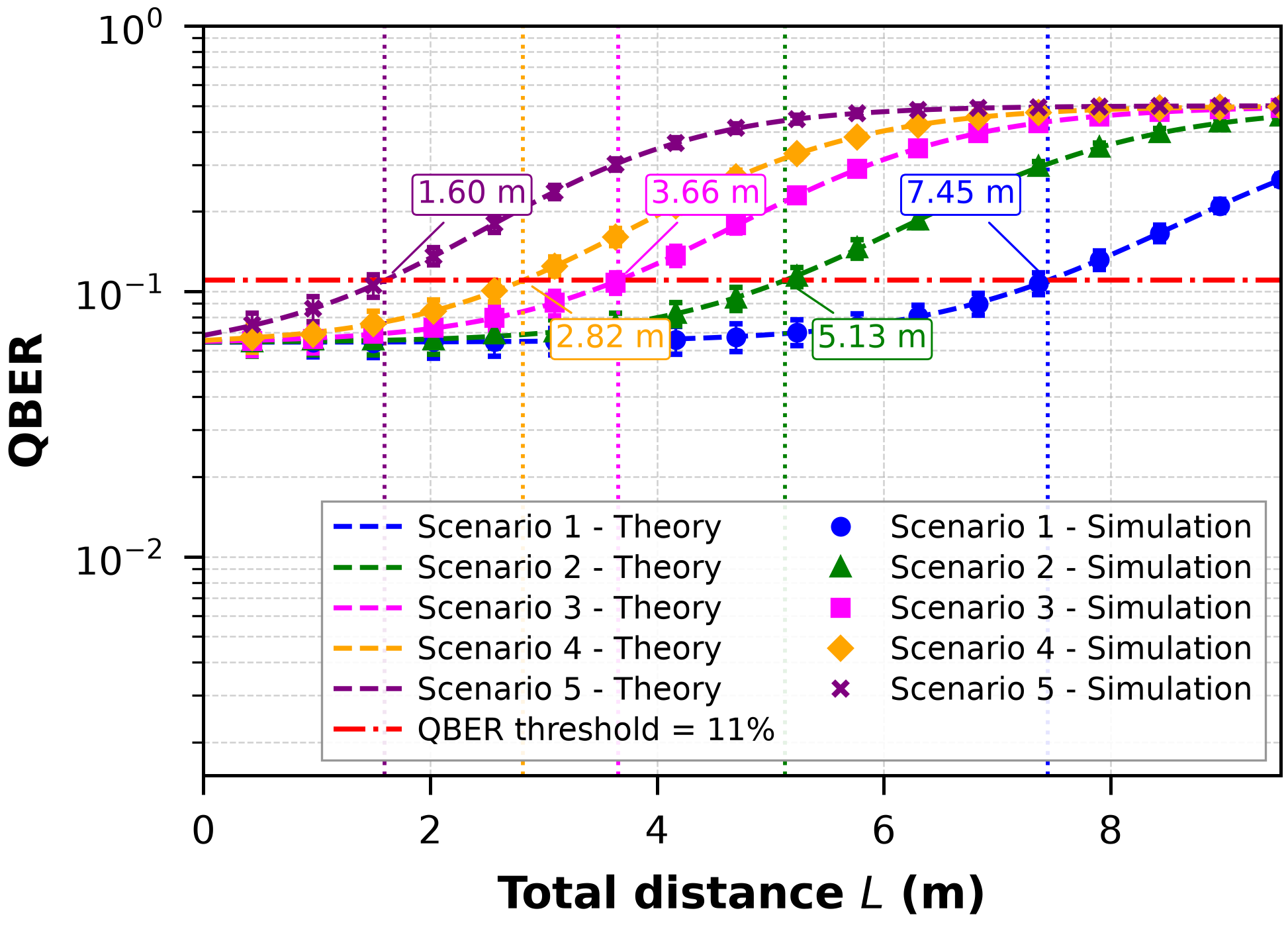}\\[-2pt]
    {\small (c)}
    \caption{\small \chg{QBER of BBM92 versus the total distance $L$ for $\beta=30^\circ$, $x=0.3L$ and $d_1=d_2=10$~cm, under the five irradiance scenarios, in (a) clear water, (b) coastal water and (c) turbid water.}}
\label{figure 2}
\end{figure}

\begin{revblock}
To assess the effect of an imperfect source, Fig.~\ref{figure 2} presents the QBER for $\beta=30^\circ$, $x=0.3L$ and $d_1=d_2=10$~cm. At short distances, the QBER starts at about $6.4\,\%$ instead of $3.3\,\%$, owing to the intrinsic error $e_\beta=[1-\sin(2\beta)]/4\simeq3.3\,\%$ of the non-maximally entangled state. In Fig.~\ref{figure 2}(a), which corresponds to clear water, the threshold is reached at $179.66$, $123.73$, $88.30$, $68.01$ and $38.58$~m for Scenarios~1 to~5. These distances decrease to $56.69$, $39.04$, $27.86$, $21.46$ and $12.17$~m in coastal water, as shown in Fig.~\ref{figure 2}(b), and to $7.45$, $5.13$, $3.66$, $2.82$ and $1.60$~m in turbid water, as shown in Fig.~\ref{figure 2}(c).

The differences with Fig.~\ref{figure 1} result from the joint action of the entanglement degree, the aperture diameters and the source position. Under Scenario~1, where background light is very weak, the lower entanglement increases the error of true coincidences while the smaller transmitter aperture increases the propagation losses, so that the threshold distances are shorter in the three water types. When the irradiance increases, the smaller receiver aperture $d_2$ collects less ambient light, which limits the background contribution to the QBER. For the most irradiated scenarios, this effect outweighs the larger losses and the non-maximality of the state, leading to longer threshold distances than those obtained in Fig.~\ref{figure 1}, e.g.\ $38.58$~m instead of $7.78$~m in clear water under Scenario~5. The aperture diameter should therefore be adapted to the illumination conditions, since large apertures are beneficial at night, whereas small apertures are preferable in daylight. Finally, the asymmetric position $x=0.3L$ makes Bob's photon travel $0.7L$, which increases the loss and depolarization on this arm. This effect is quantified in Section~\ref{sec_position}.
\end{revblock}

\subsection{\chg{Optimization of the source position}}
\label{sec_position}

\begin{figure}[!htbp]
    \centering
    \includegraphics[width=1.0\linewidth]{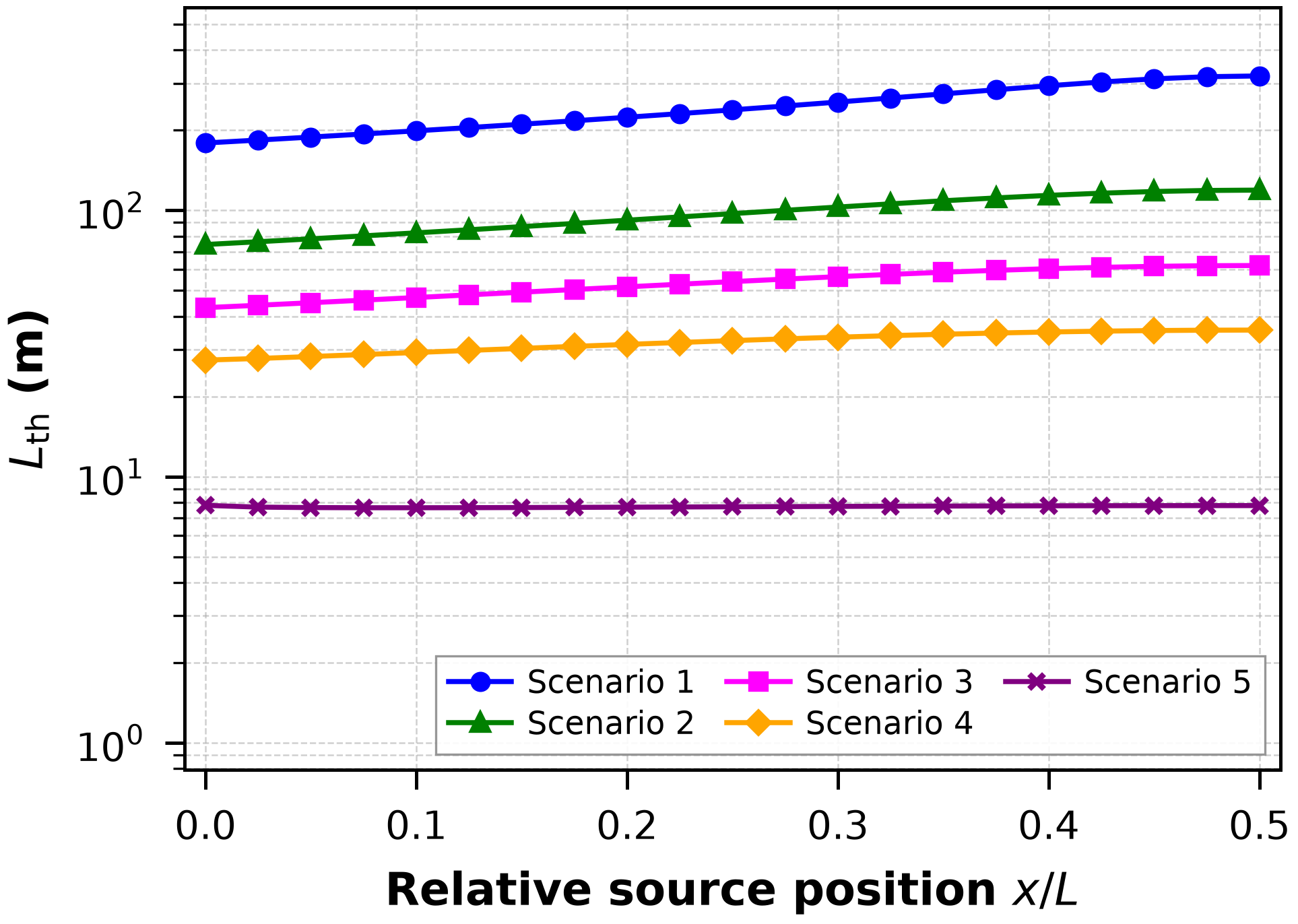}\\[-2pt]
    {\small (a)}\\[5pt]
    \includegraphics[width=1.0\linewidth]{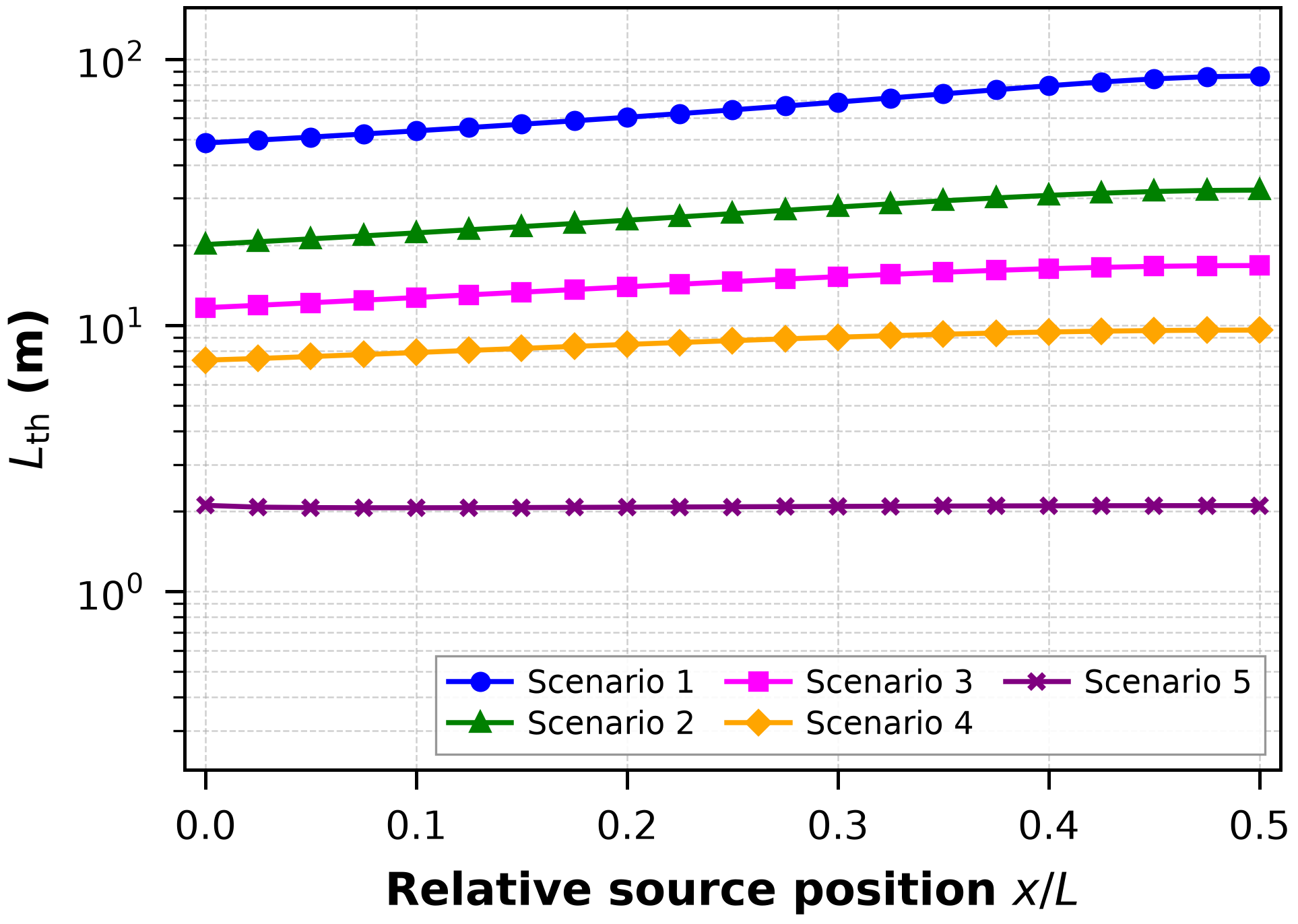}\\[-2pt]
    {\small (b)}\\[5pt]
    \includegraphics[width=1.0\linewidth]{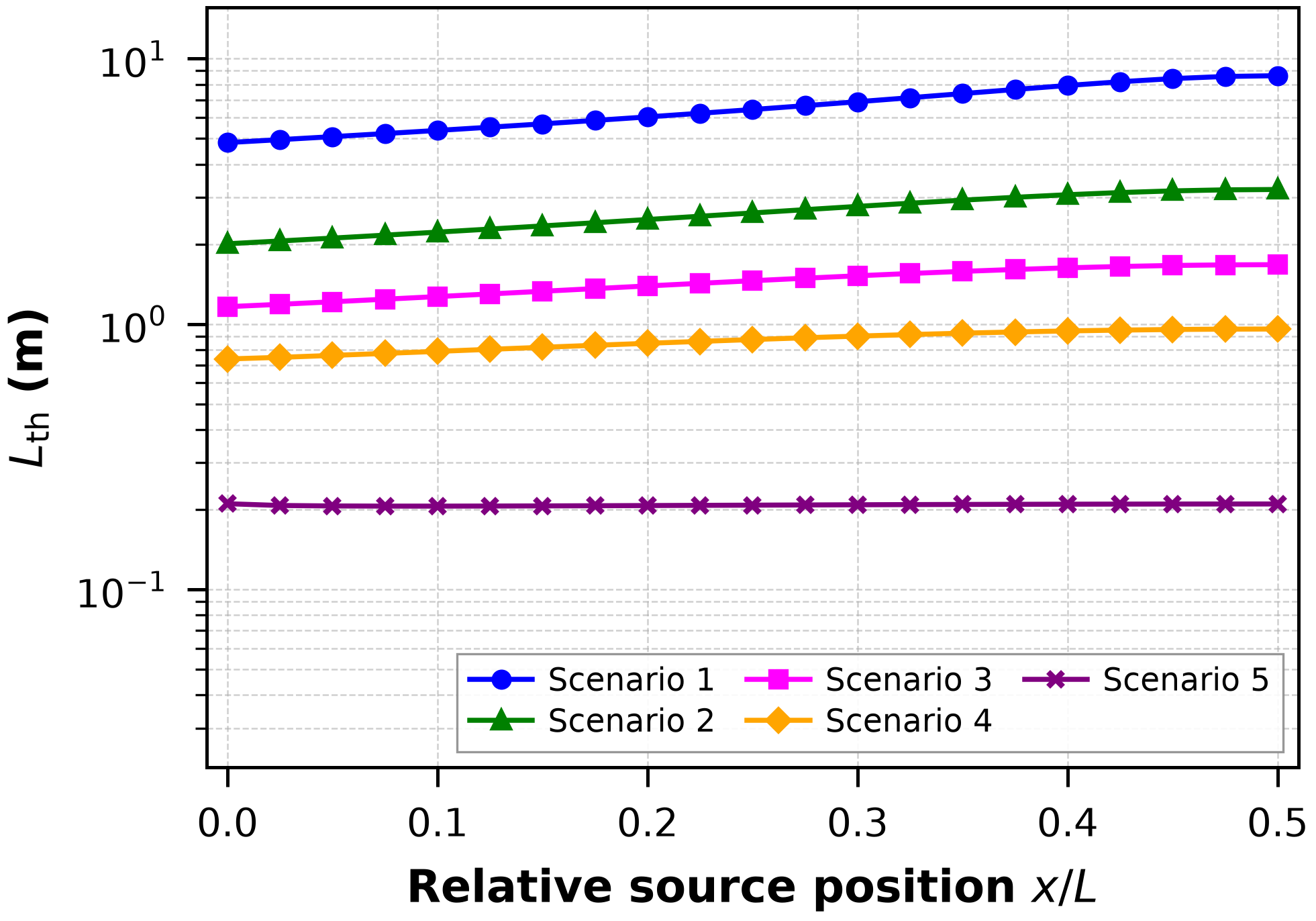}\\[-2pt]
    {\small (c)}
    \caption{\small \chg{Threshold distance $L_{\mathrm{th}}$ versus the relative source position $x/L$, for $\beta=45^\circ$ and $d_1=d_2=30$~cm, under the five irradiance scenarios, in (a) clear water, (b) coastal water and (c) turbid water. By symmetry, the curves for $x/L\in[0.5,1]$ are mirror images.}}
\label{fig_position}
\end{figure}

\begin{figure}[!htbp]
    \centering
    \includegraphics[width=1.0\linewidth]{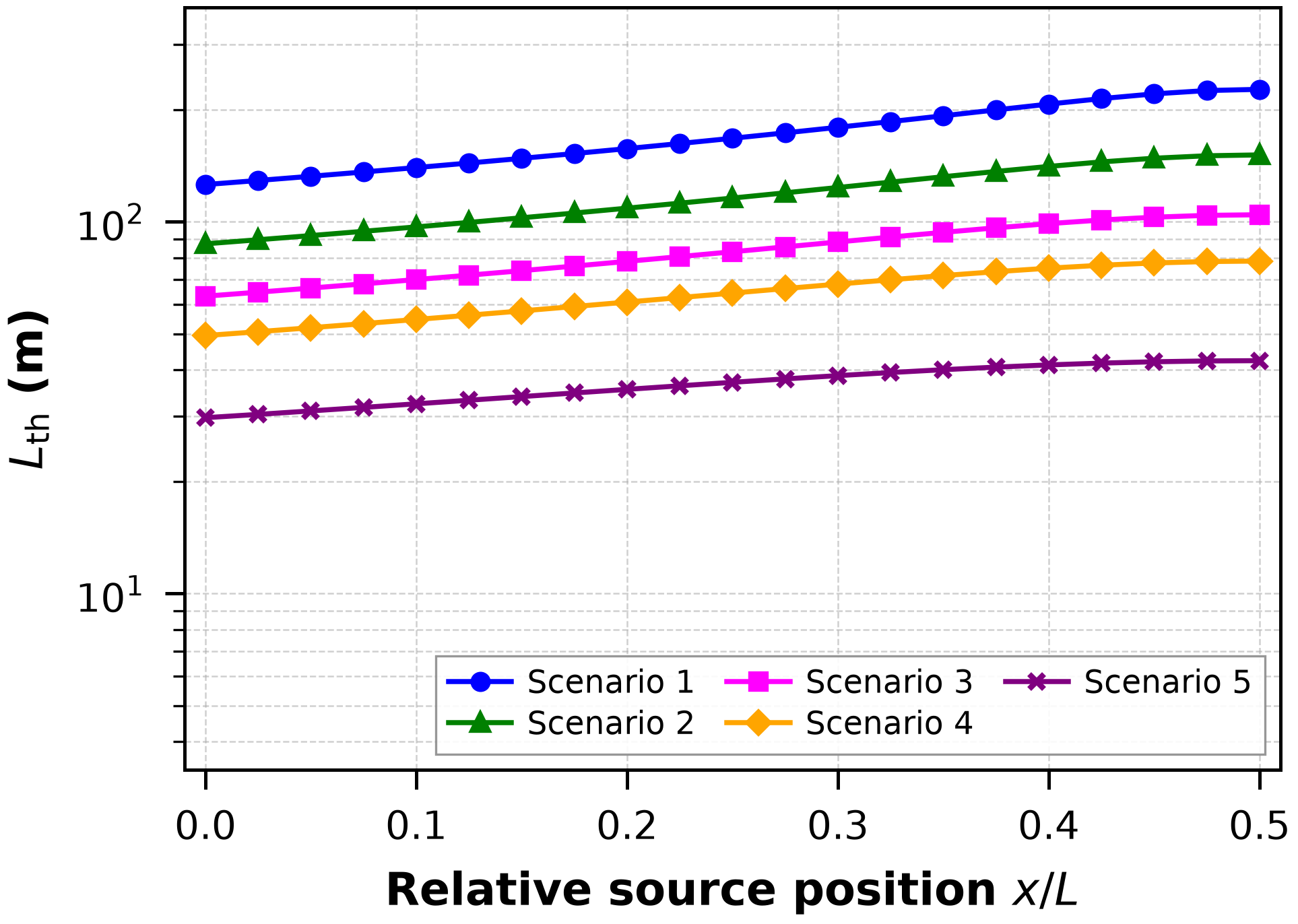}\\[-2pt]
    {\small (a)}\\[5pt]
    \includegraphics[width=1.0\linewidth]{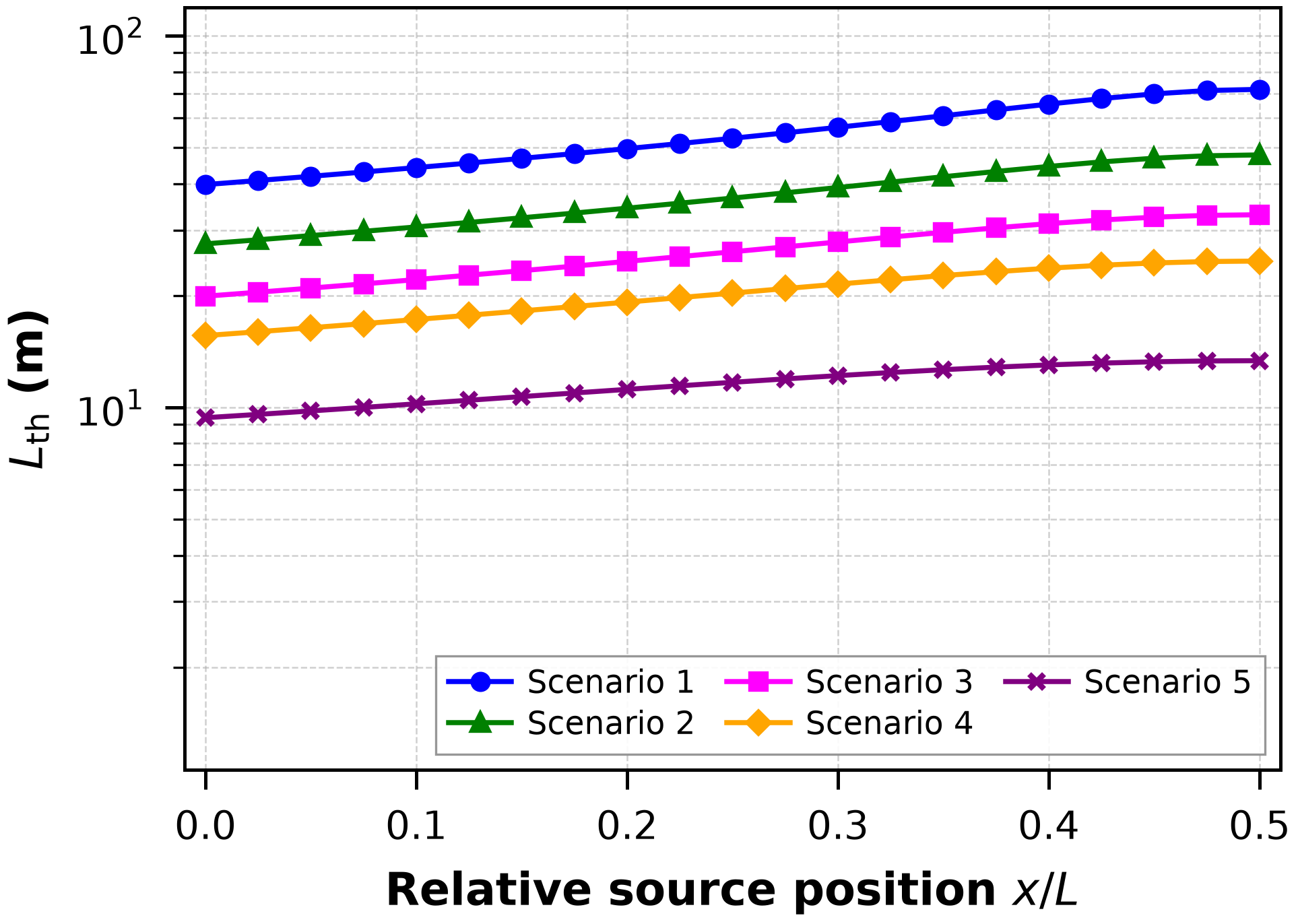}\\[-2pt]
    {\small (b)}\\[5pt]
    \includegraphics[width=1.0\linewidth]{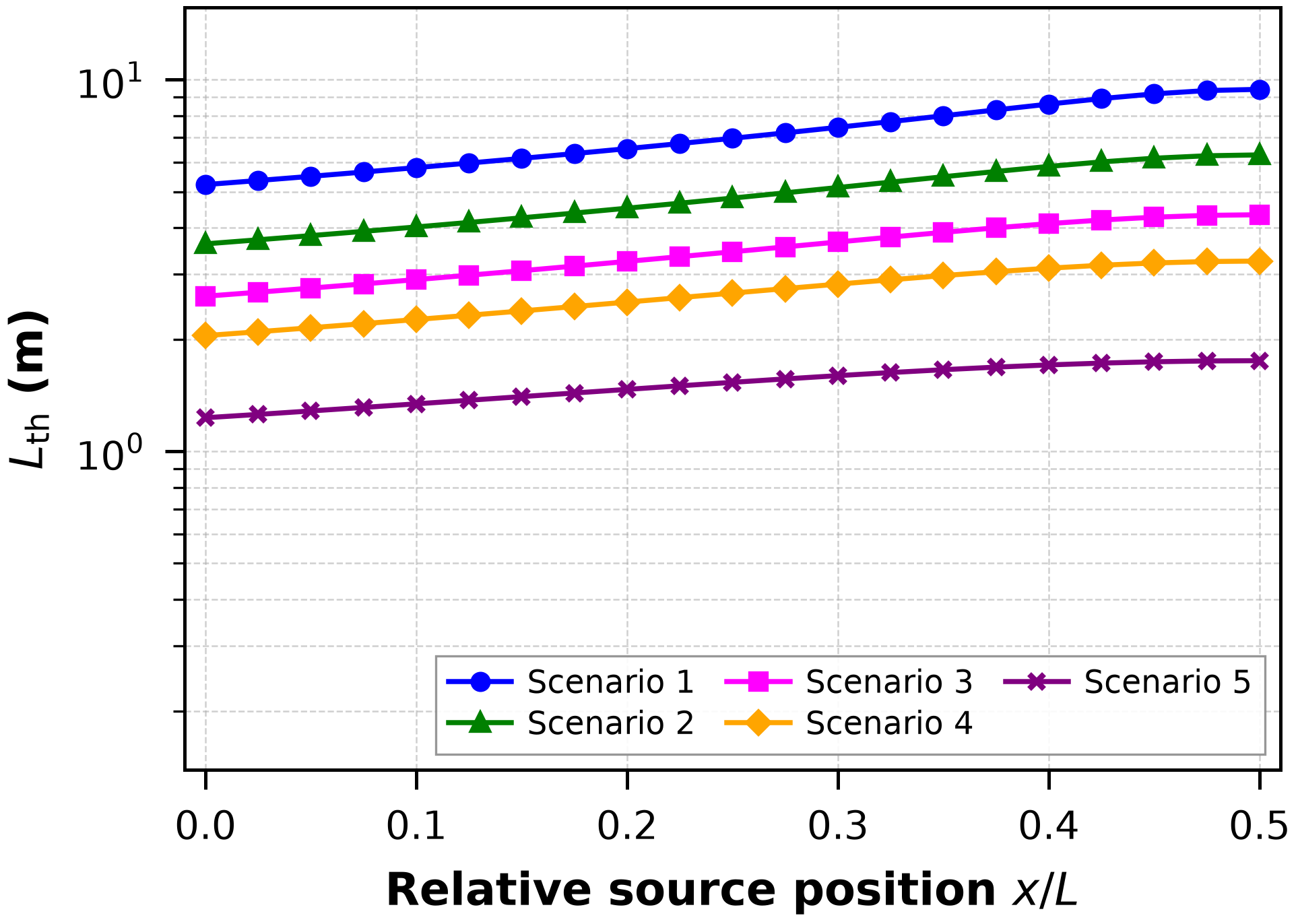}\\[-2pt]
    {\small (c)}
    \caption{\small \chg{Threshold distance $L_{\mathrm{th}}$ versus the relative source position $x/L$, for $\beta=30^\circ$ and $d_1=d_2=10$~cm, under the five irradiance scenarios, in (a) clear water, (b) coastal water and (c) turbid water.}}
\label{fig_position30}
\end{figure}

\begin{figure}[!htbp]
    \centering
    \includegraphics[width=1.0\linewidth]{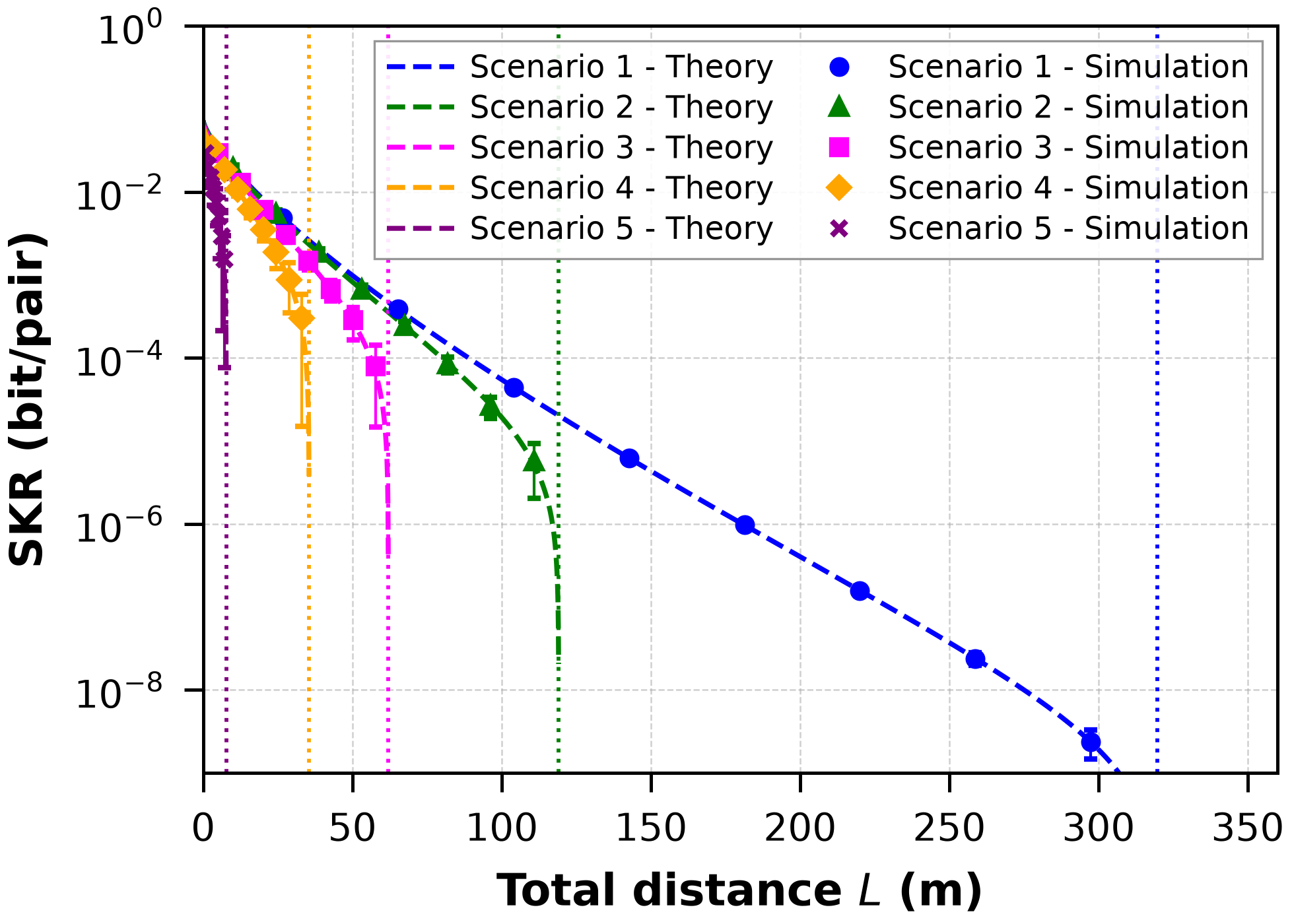}\\[-2pt]
    {\small (a)}\\[5pt]
    \includegraphics[width=1.0\linewidth]{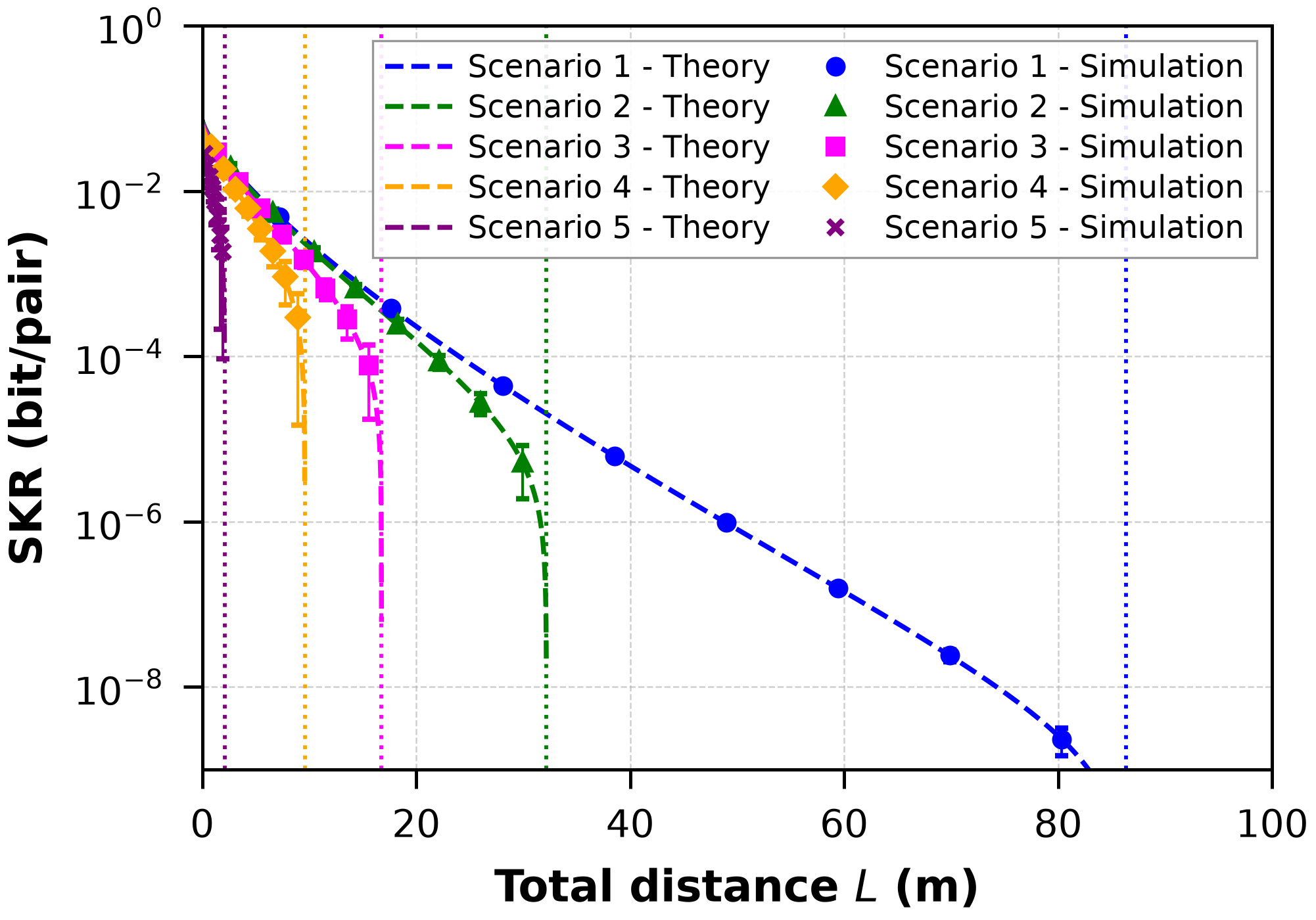}\\[-2pt]
    {\small (b)}\\[5pt]
    \includegraphics[width=1.0\linewidth]{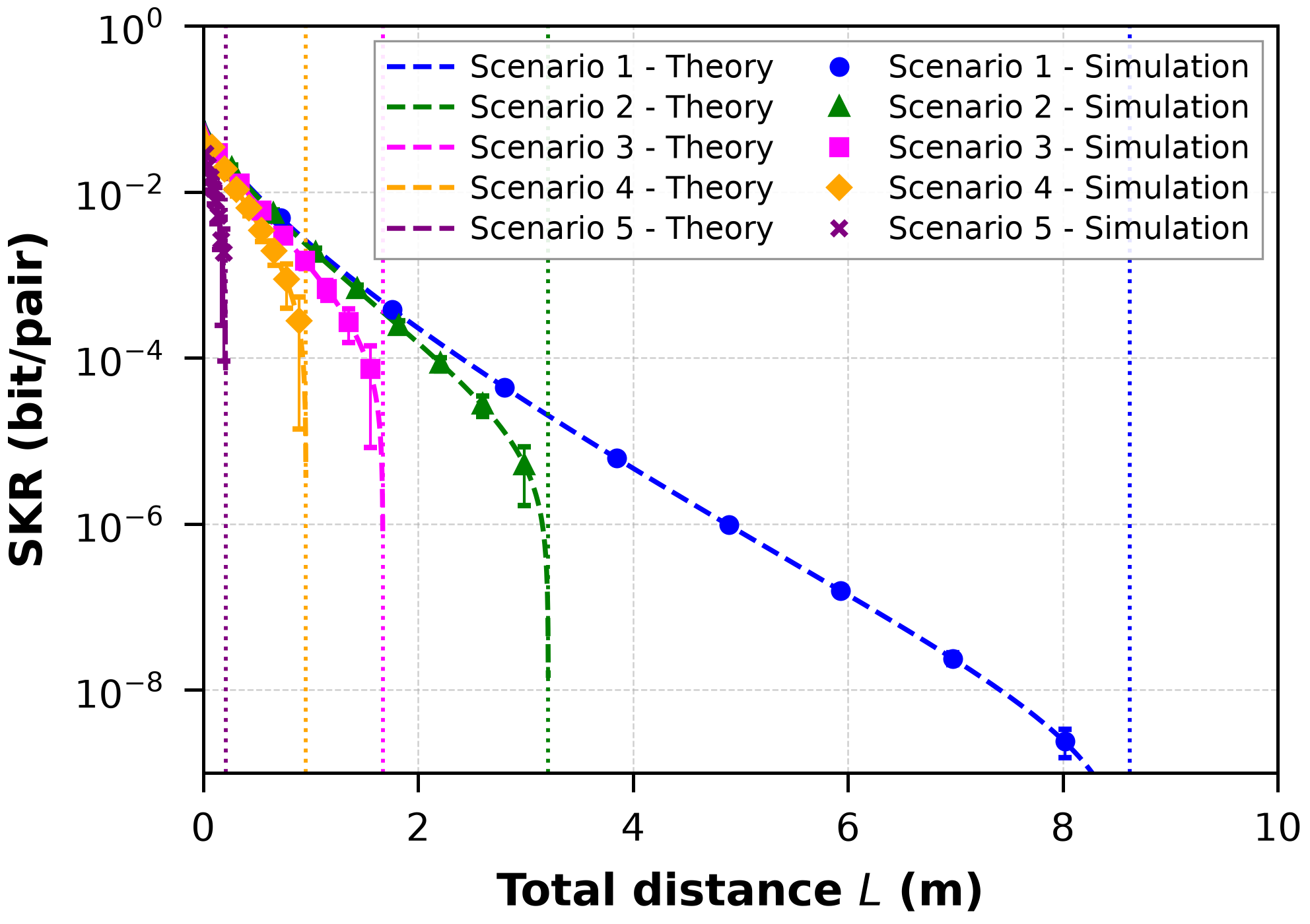}\\[-2pt]
    {\small (c)}
    \caption{\small \chg{SKR of BBM92 (bit per generated pair) versus the total distance $L$ for $\beta=45^\circ$, $x=0.5L$ and $d_1=d_2=30$~cm, under the five irradiance scenarios, in (a) clear water, (b) coastal water and (c) turbid water. Vertical dotted lines indicate the QBER threshold distances.}}
\label{fig_skr45}
\end{figure}

\begin{figure}[!htbp]
    \centering
    \includegraphics[width=1.0\linewidth]{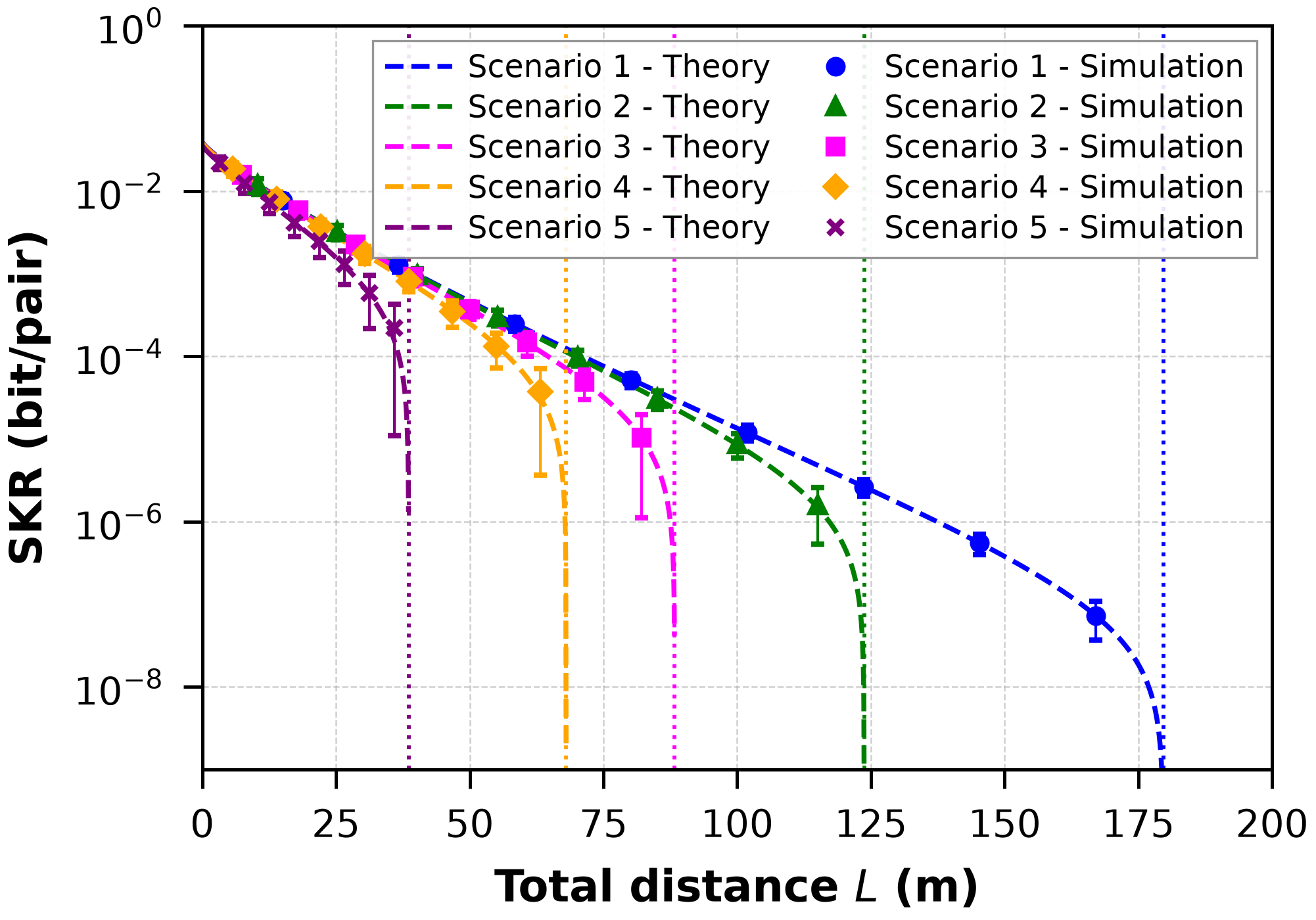}\\[-2pt]
    {\small (a)}\\[5pt]
    \includegraphics[width=1.0\linewidth]{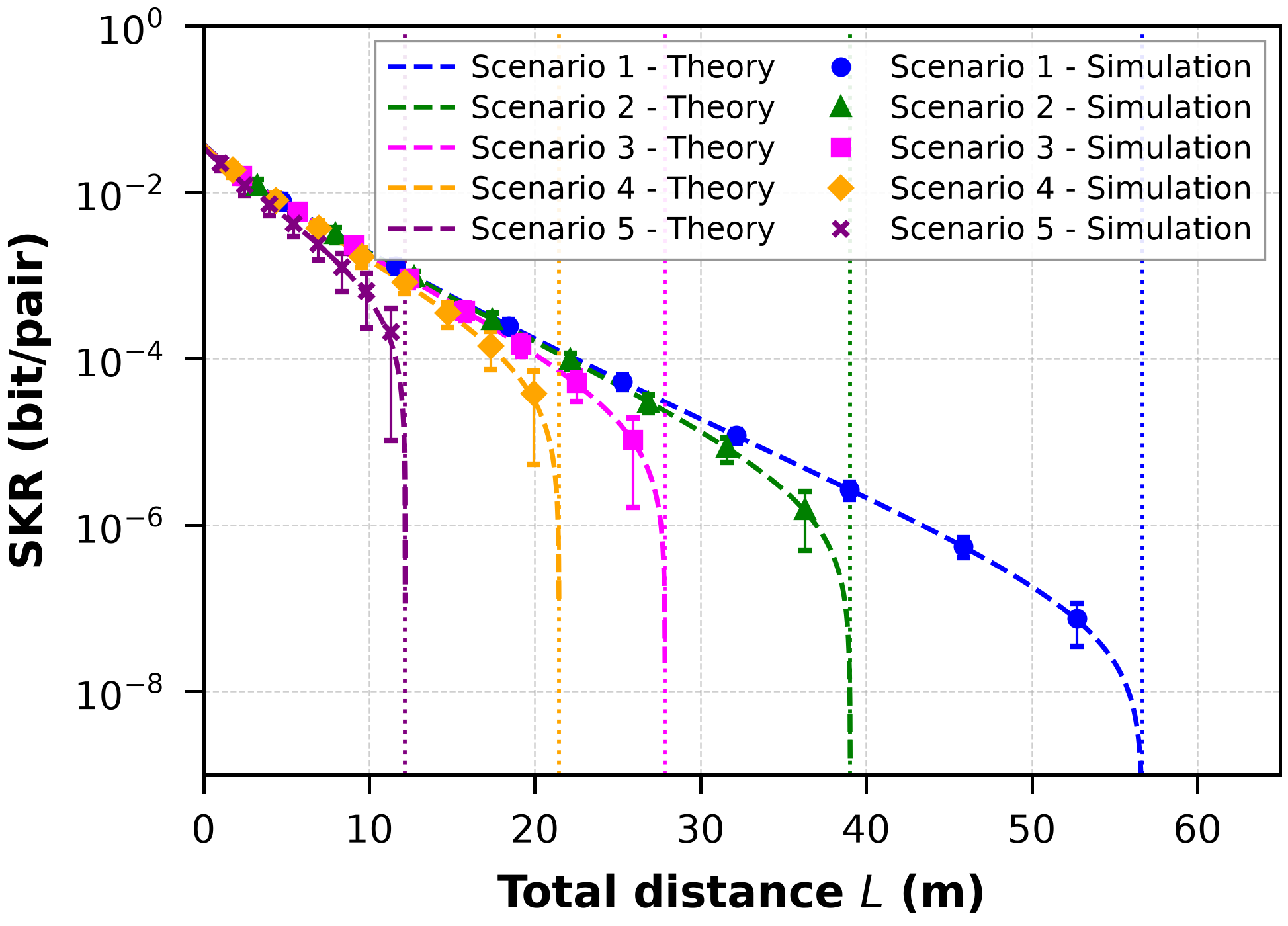}\\[-2pt]
    {\small (b)}\\[5pt]
    \includegraphics[width=1.0\linewidth]{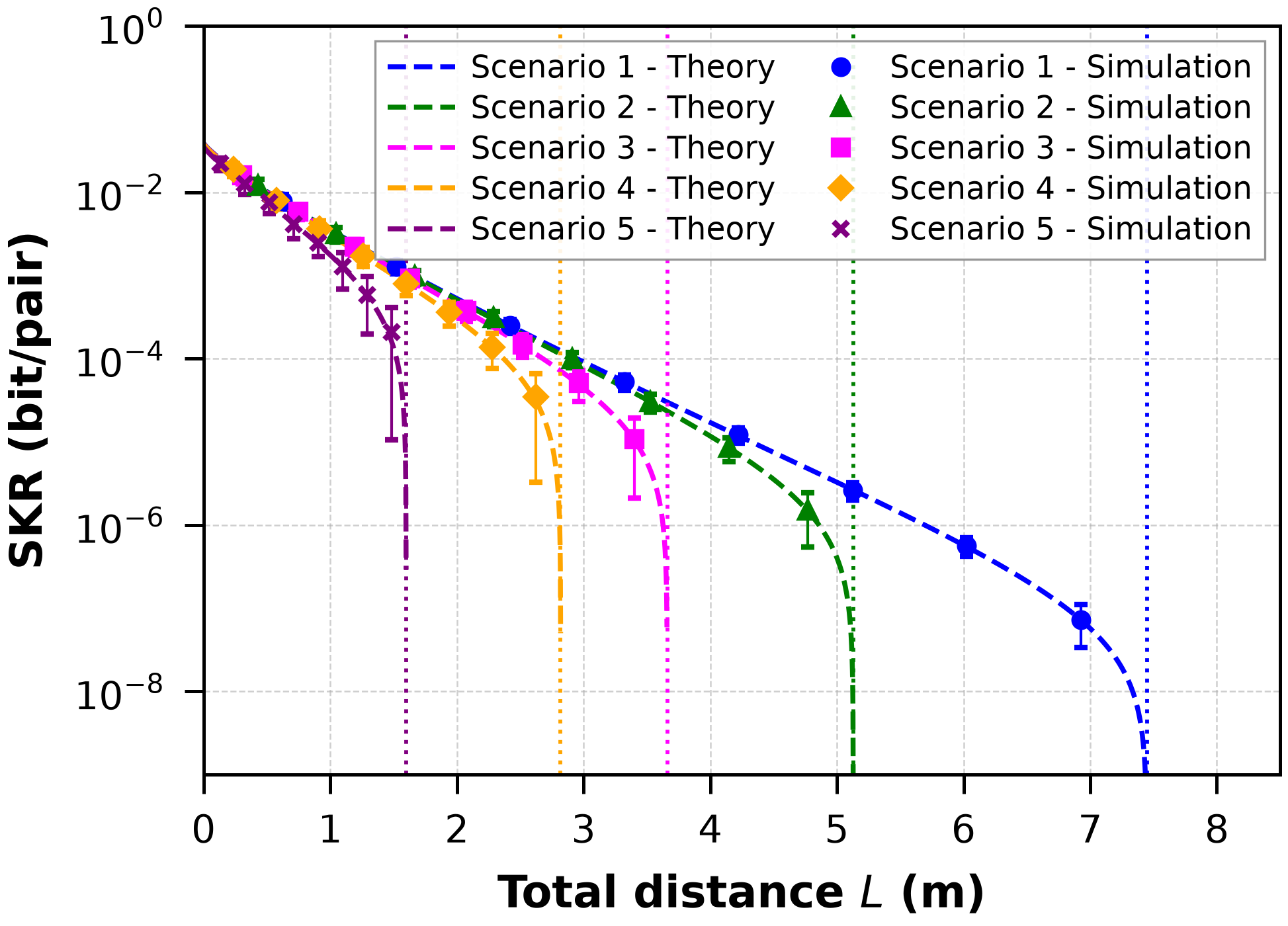}\\[-2pt]
    {\small (c)}
    \caption{\small \chg{SKR of BBM92 (bit per generated pair) versus the total distance $L$ for $\beta=30^\circ$, $x=0.3L$ and $d_1=d_2=10$~cm, under the five irradiance scenarios, in (a) clear water, (b) coastal water and (c) turbid water.}}
\label{figure 3}
\end{figure}

\begin{revblock}
The previous configurations used a centred source for $\beta=45^\circ$ and an asymmetric source for $\beta=30^\circ$, which raises the question of the optimal position of the source along the link. Figs.~\ref{fig_position} and \ref{fig_position30} show $L_{\mathrm{th}}$ as a function of $x/L$ for the two reference configurations. For all scenarios and water types, the threshold distance increases monotonically when $x/L$ goes from $0$ to $0.5$, and the symmetric placement $x=L/2$ is optimal. In Fig.~\ref{fig_position}(a), which corresponds to $\beta=45^\circ$ in clear water, $L_{\mathrm{th}}$ increases under Scenario~1 from $179.08$~m for a source co-located with Alice ($x\to0$) to $198.93$~m at $x=0.1L$, $223.70$~m at $x=0.2L$ and $319.78$~m at $x=0.5L$. The same trend holds in coastal water, with $48.35$, $60.40$ and $86.36$~m for $x\to0$, $0.2L$ and $0.5L$, and in turbid water, with $4.83$, $6.03$ and $8.62$~m, as shown in Figs.~\ref{fig_position}(b) and \ref{fig_position}(c). In Fig.~\ref{fig_position30}, which corresponds to $\beta=30^\circ$ and $d_1=d_2=10$~cm, $L_{\mathrm{th}}=126.03$, $157.44$, $179.66$ and $227.30$~m in clear water under Scenario~1 for $x\to0$, $0.2L$, $0.3L$ and $0.5L$.

The mechanism behind this behaviour is explained by Eq.~\eqref{QBER_BBM92_non_max}. With identical receivers in a homogeneous medium, the exchange $x\leftrightarrow L-x$ leaves the system invariant, so that $x=L/2$ is a stationary point. Depolarization is minimal at $x=L/2$, since $\ln\kappa_X$ is a concave function of $\ell_X$. The dominant effect, however, concerns the fraction of true coincidences. When the source is close to Alice, her photon is detected with high probability, and each of these detections can form an accidental coincidence with a noise click at Bob's receiver, whose own photon is strongly attenuated. The signal-noise term $\eta_A(1-\eta_B)y_0$ of Eq.~\eqref{eq_pfalse} then dominates the true coincidences $\eta_A\eta_B$. Balancing the two arms minimizes $\mathbb{P}_{\mathrm{false}}/\mathbb{P}_{\mathrm{true}}=(1+y_0u_A)(1+y_0u_B)-1$, with $u_X=1/\eta_X-1$, and thus the QBER. Note that, for $T>0$, the joint transmittance Eq.~\eqref{eq_joint_transmittance} is actually slightly lower at $x=L/2$ than for an asymmetric placement, so that the benefit of the centred source therefore comes from the reduction of accidental coincidences and not from a higher transmission.

This behaviour can be derived analytically. With $p_{\mathrm{pair}}=1$, Eqs.~\eqref{Proba coincidence} and \eqref{eq_ptrue} give
\begin{equation}
\label{eq_ratio_true}
\frac{\mathbb{P}_{\mathrm{true}}}{\mathbb{P}_{\mathrm{coinc}}}=\frac{\eta_A\eta_B}{c_Ac_B}=\frac{1}{(1+y_0u_A)(1+y_0u_B)},\quad u_X=\frac{1}{\eta_X}-1,
\end{equation}
so that, from the second form of Eq.~\eqref{QBER_BBM92_non_max}, the QBER decreases when $e_{\mathrm{sig}}$ decreases and when $F(x)=\ln(1+y_0u_A)+\ln(1+y_0u_B)$ decreases. For the depolarization term, Eqs.~\eqref{eq_e_beta_dep} and \eqref{proba canal+detector} show that $e_{\mathrm{sig}}$ decreases when $r$ increases. With $\kappa_X=\frac13(4e^{-\gamma_{\mathrm{dep}}\ell_X}-1)$,
\begin{equation}
\frac{\mathrm{d}^2\ln\kappa_X}{\mathrm{d}\ell_X^2}=-\frac{4\gamma_{\mathrm{dep}}^2e^{\gamma_{\mathrm{dep}}\ell_X}}{\big(4-e^{\gamma_{\mathrm{dep}}\ell_X}\big)^2}<0,
\end{equation}
so that $\ln r=\ln\kappa_A(x)+\ln\kappa_B(L-x)$ is a concave function of $x$, symmetric with respect to $L/2$, and is therefore maximal at $x=L/2$. For the coincidence term, with $\eta_X=\eta_0\,e^{-g(\ell_X)}$, $\eta_0=\eta_{\mathrm{Alice}}=\eta_{\mathrm{Bob}}$ and $g(\ell)=\alpha(d_1/\theta_{\mathrm{rad}})^T\ell^{1-T}$, the function $f(\ell)=\ln[1+y_0(e^{g(\ell)}/\eta_0-1)]$ satisfies $F(x)=f(x)+f(L-x)$, and $F'(L/2)=0$ by symmetry. For $T=0$, $g(\ell)=\alpha\ell$ and
\begin{equation}
f'(\ell)=\alpha\,\frac{w(\ell)}{1-y_0+w(\ell)},\qquad w(\ell)=\frac{y_0\,e^{\alpha\ell}}{\eta_0},
\end{equation}
which is an increasing function of $\ell$. Hence, $f$ is convex and $F$ is minimal at $x=L/2$. In this case, $\mathbb{P}_{\mathrm{true}}=\eta_0^2e^{-\alpha L}$ does not depend on $x$, and the optimality of the centred source entirely results from the minimization of the accidental coincidences. For $T>0$, $g$ is concave, so that convexity is no longer guaranteed analytically, and the global minimum of the QBER at $x=L/2$ is then verified numerically, as shown in Figs.~\ref{fig_position} and \ref{fig_position30}.

The dependence on $x$ becomes weaker when the irradiance increases, because the range is then short and both arms have comparable transmittances. For instance, for $\beta=45^\circ$ under Scenario~5 in clear water, $L_{\mathrm{th}}=7.68$~m at $x=0.2L$ and $7.78$~m at $x=0.5L$. It remains significant for $\beta=30^\circ$ and $d_1=d_2=10$~cm, where the range is longer, since $L_{\mathrm{th}}$ increases from $29.73$~m at $x\to0$ to $42.33$~m at $x=0.5L$ under Scenario~5 in clear water. A centred source is therefore the recommended configuration, while an asymmetric placement, which may be imposed by practical constraints such as a source hosted on one of the platforms, has a cost that can be read in Figs.~\ref{fig_position} and \ref{fig_position30}.
\end{revblock}

\begin{figure}[!htbp]
    \centering
    \includegraphics[width=1.0\linewidth]{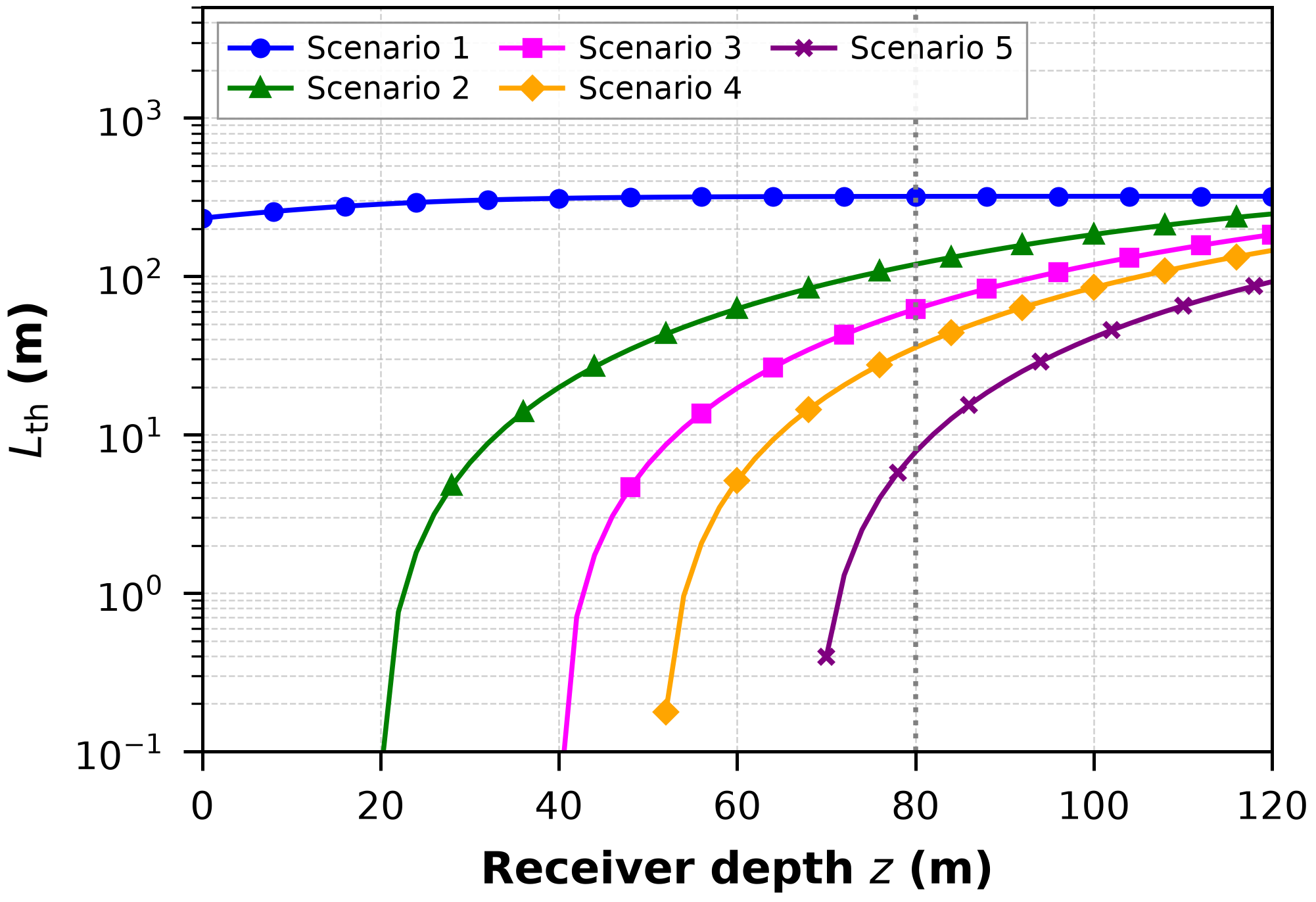}\\[-2pt]
    {\small (a)}\\[5pt]
    \includegraphics[width=1.0\linewidth]{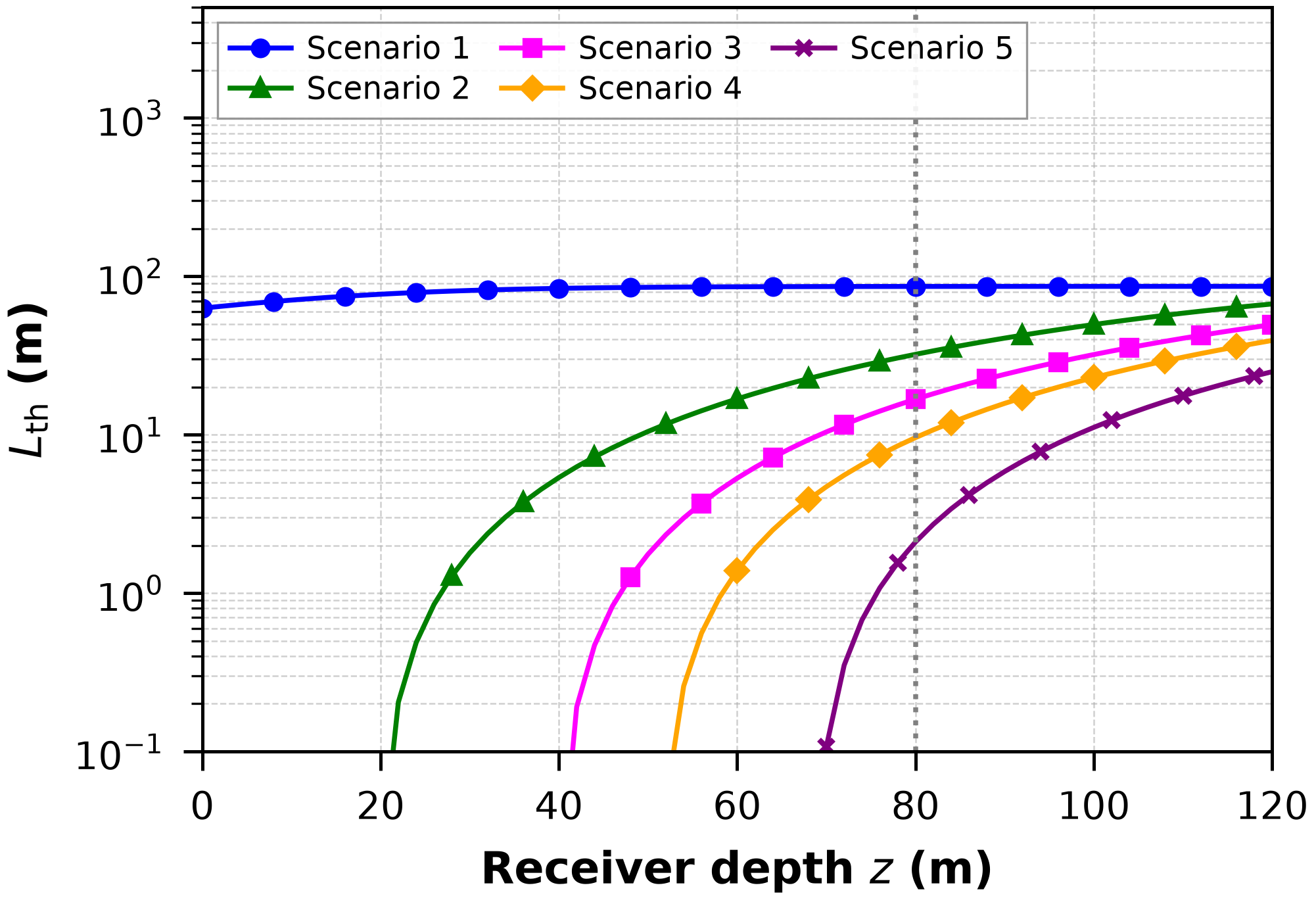}\\[-2pt]
    {\small (b)}
    \caption{\small \chg{Threshold distance $L_{\mathrm{th}}$ versus the receiver depth $z$ for $\beta=45^\circ$, $x=0.5L$ and $d_1=d_2=30$~cm, under the five irradiance scenarios, in (a) clear water and (b) coastal water. The vertical dotted line indicates the reference depth $z=80$~m.}}
\label{fig_depth}
\end{figure}

\subsection{SKR \chg{analysis}}

\begin{revblock}
Once the conditions for a secure key are identified, the amount of key that can be generated is given by the SKR. Figs.~\ref{fig_skr45} and \ref{figure 3} show the SKR for the two reference configurations. The theoretical results agree with the Monte Carlo simulations. In Fig.~\ref{fig_skr45}, which corresponds to $\beta=45^\circ$, the SKR reaches about $7.2\times10^{-2}$ bit per pair at short distances in clear water, and $7.1\times10^{-2}$ and $6.5\times10^{-2}$ in coastal and turbid water, where losses are already significant over the first centimetres. In Fig.~\ref{figure 3}, which corresponds to $\beta=30^\circ$, it starts at about $3.9\times10^{-2}$ bit per pair, since the intrinsic error $e_\beta$ roughly halves the secret fraction. At short distances, the five scenarios give similar values, and the curves then separate as the distance and the irradiance increase. This evolution results from the decrease of the coincidence probability, which makes the SKR decrease exponentially, and from the increase of the QBER, which makes the secret fraction collapse when approaching the threshold. Since $f_{\mathrm{EC}}\simeq1$, the SKR vanishes at the threshold distances obtained in Figs.~\ref{figure 1} and \ref{figure 2}. The region in which a secret key can be generated is widest in clear water and shrinks in coastal and turbid water, which directly reflects the influence of attenuation on the protocol performance. With a pair rate of a few MHz, an SKR of $10^{-4}$ bit per pair corresponds to a few hundred secret bits per second, which is relevant for the periodic renewal of symmetric keys on an underwater link.
\end{revblock}

\subsection{\chg{Sensitivity to the depolarization coefficient}}
\label{sec_gamma}
\begin{revblock}
The previous results show that depolarization has a weak effect compared with attenuation and background light. Since this conclusion relies on the value of $\gamma_{\mathrm{dep}}$, its robustness must be checked. The values of $\gamma_{\mathrm{dep}}$ used in this work are obtained from a measurement at $810$~nm through the wavelength scaling factor $10^{-2}$ of Eq.~\eqref{eq_gamma_530}, which is based on water absorption only. Since this scaling is approximate, the true coefficients at $530$~nm may be larger. To check whether this uncertainty can change the conclusions, all coefficients are multiplied by a factor $k$, i.e.\ $\gamma_{\mathrm{dep}}\to k\,\gamma_{\mathrm{dep}}$, and the threshold distance is recomputed under Scenario~1, where depolarization has the largest relative weight since background noise is minimal. The values of $k$ have a direct physical meaning, given as follows.
\begin{itemize}
\item $k=1$ gives the nominal values of Table~\ref{Table1}, obtained with the wavelength scaling.
\item $k=10$ corresponds to a scaling factor of $10^{-1}$ instead of $10^{-2}$.
\item $k=10^2$ corresponds to no wavelength scaling at all, i.e.\ the coefficient measured at $810$~nm is used directly at $530$~nm. For turbid water, $k\,\gamma_{\mathrm{dep}}=7.62\times10^{-4}$~m$^{-1}$, which is the value of Eq.~\eqref{eq_gamma_810}.
\item $k=10^3$ corresponds to a coefficient ten times larger than the value measured at $810$~nm, which is an extreme case since water absorbs less at $530$~nm than at $810$~nm.
\item $k=0$ corresponds to a channel without depolarization, given as a reference.
\end{itemize}
Table~\ref{tab_gamma} gives the results for the two reference configurations. For instance, in clear water with $\beta=45^\circ$, $L_{\mathrm{th}}=319.78$~m with the nominal coefficient, $317.01$~m for $k=10$, and still $283.52$~m ($-11\,\%$) for $k=10^2$, i.e.\ even if the wavelength scaling is completely removed. Only for the extreme case $k=10^3$ does depolarization become the limiting mechanism in clear and coastal water. In turbid water, where the range is limited to a few metres by attenuation, depolarization remains negligible even for $k=10^3$. The conclusions of this work therefore do not depend on the approximate wavelength scaling of $\gamma_{\mathrm{dep}}$, and the dominant limitations remain attenuation and background light. A direct measurement of $\gamma_{\mathrm{dep}}$ at $530$~nm, accounting for the phase function and the particle size distribution, would nevertheless refine the model.

\begin{table*}[!t]
\centering
\caption{Threshold distance $L_{\mathrm{th}}$ (m) under Scenario~1 when all depolarization coefficients are multiplied by a factor $k$, to test the robustness of the results to the wavelength scaling of $\gamma_{\mathrm{dep}}$ (Eq.~\eqref{eq_gamma_530}). The nominal values ($k=1$) are in bold.}
\label{tab_gamma}
\renewcommand{\arraystretch}{1.25}
\begin{tabular}{@{}cl|*{3}{>{\centering\arraybackslash}p{1.3cm}}|*{3}{>{\centering\arraybackslash}p{1.3cm}}@{}}
\toprule
 & & \multicolumn{3}{c|}{(i) $\beta=45^\circ$, $x=0.5L$, $d_1=d_2=30$~cm} & \multicolumn{3}{c}{(ii) $\beta=30^\circ$, $x=0.3L$, $d_1=d_2=10$~cm}\\
$k$ & Meaning & Clear & Coastal & Turbid & Clear & Coastal & Turbid\\
\midrule
$0$      & No depolarization                & 320.09 & 86.39 & 8.62 & 179.80 & 56.72 & 7.45\\
$1$      & Nominal values (Table~\ref{Table1}) & \textbf{319.78} & \textbf{86.36} & \textbf{8.62} & \textbf{179.66} & \textbf{56.69} & \textbf{7.45}\\
$10$     & Scaling factor $10^{-1}$ instead of $10^{-2}$ & 317.01 & 86.05 & 8.62 & 178.42 & 56.51 & 7.44\\
$10^{2}$ & No wavelength scaling (810~nm value) & 283.52 & 82.73 & 8.55 & 163.38 & 54.45 & 7.37\\
$10^{3}$ & $10\times$ the 810~nm value (extreme) & 55.55  & 35.49 & 7.81 & 34.29  & 22.07 & 6.49\\
\bottomrule
\end{tabular}
\end{table*}
\end{revblock}

\subsection{\chg{Daytime background light, receiver depth and detector constraints}}
\label{sec_daylight}

\begin{revblock}
Background light was identified above as one of the two dominant limitations. Its influence is now analysed as a function of the receiver depth, together with its practical consequences for the detectors. The background term of Eq.~\eqref{eq_nB} decreases as $e^{-K_\infty z}$, so that daytime operation strongly depends on the receiver depth. Fig.~\ref{fig_depth} shows $L_{\mathrm{th}}$ as a function of $z$. In Fig.~\ref{fig_depth}(a), which corresponds to clear water, the range under Scenario~1 is limited by attenuation and dark counts and is almost independent of depth, with $233$~m at the surface and $320$~m at $z\geq80$~m. Under daylight, in contrast, a minimum depth is required for any secret key to be generated, about $20$~m for Scenario~2 and $70$~m for Scenario~5 in both water types, as shown in Figs.~\ref{fig_depth}(a) and \ref{fig_depth}(b). At shallower depths, background clicks dominate the coincidences and the QBER exceeds $11\,\%$ at any distance. The protocol is then not operable, which is the regime expected in shallow coastal waters under intense zenith sunlight. Below these depths, the range increases rapidly with $z$, reaching for instance $41.21$~m at $z=100$~m under Scenario~5 in clear water.

At $z=80$~m and $d_2=30$~cm, the model gives $n_B\simeq7.9\times10^{-8}$ background photoelectrons per $200$~ps window and receiver under Scenario~1, and $n_B\simeq3.9\times10^{-2}$ under Scenario~5, compared with $\overline{n}_{\mathrm{dc}}\simeq9.6\times10^{-6}$. These values have three practical implications. First, temporal gating is essential. With a $200$~ps gate every $40$~ns, the noise click probability under Scenario~5 is $y_0\simeq3.9\times10^{-2}$ per window, i.e.\ about $10^6$ clicks/s per receiver ($\sim2.4\times10^5$ per detector), well below the saturation of typical single-photon avalanche diodes (SPADs), whereas the same photon flux on a free-running detector would correspond to about $2\times10^8$ photons/s per receiver and would saturate it. Second, narrow spectral filtering is required, since $n_B$ scales linearly with $\Delta\lambda$. Third, the angular acceptance must be restricted, since $n_B$ scales with the product $E\,\Omega_{\mathrm{eff}}$, and the calibration factor $F_{\mathrm{eq}}\simeq10^{-5}$ applied to the irradiance is equivalent to using the true surface irradiance with a receiver full field of view of $2\arcsin\sqrt{F_{\mathrm{eq}}}\simeq6.3$~mrad instead of $180^\circ$ in Eq.~\eqref{eq_omega}. A wide-field receiver exposed to the unattenuated sunlight would therefore be saturated at shallow depths, which confirms that narrow-field optics, narrowband filtering and gated detection are prerequisites for daytime underwater EBQKD. Afterpulsing, detector dead time and latching are not included in the present model, and they would further penalize the most irradiated scenarios and call for detectors with fast active quenching.
\end{revblock}

\FloatBarrier
\section{Conclusion}
\label{Section6}
\begin{revblock}
In this work, we developed a quantum model of the BBM92 protocol over a non-turbulent underwater channel, in which the source emits non-maximally entangled photon pairs and the propagation medium is described by Kraus operators. Each arm combines a polarization-independent photon-loss channel, which reduces the coincidence probabilities without altering the polarization state of the detected pairs, and a depolarizing channel, which degrades their polarization correlations. We showed that these two channels commute, so that the results do not depend on an arbitrary ordering of the two processes. This model provides closed-form expressions of the output density matrix, the quantum correlations, the QBER and the SKR. All analytical results were confirmed by Monte Carlo simulations.

The quantum description separates the role of each physical mechanism. The degree of entanglement sets the initial level of the correlations, and a secret key can only be extracted when the concurrence of the source exceeds $0.67$, whatever the water type. Depolarization, in contrast, remains a weak effect over the distances considered, even when its coefficient is increased by two orders of magnitude, so that attenuation and background light are the dominant limitations. For a maximally entangled source placed at the centre of the link, the QBER remains below $11\,\%$ up to about $320$~m in clear water, $86$~m in coastal water and $8.6$~m in turbid water at night. The centred position is optimal because it minimizes the accidental coincidences between a detected photon and a noise click. Under daylight, the range decreases strongly, and operation requires a sufficient depth, gated detection, narrowband filtering and narrow-field receivers, while smaller receiver apertures become beneficial.

Since these results have been obtained for a non-turbulent channel, they should be regarded as upper bounds on the achievable range. Oceanic turbulence, induced by temperature and salinity fluctuations, makes the received power a random variable and modifies the detection model itself \cite{raouf2022performance, paglierani2023primer}. Extending the present Kraus-operator framework to turbulent channels is therefore the main perspective of this work.
\end{revblock}

\section{Acknowledgments}
The authors sincerely thank the Brittany Region for its financial support and for funding this research.

\section{\chg{Author contributions}}
\chg{\textbf{Nour Rizk}: Conceptualization, Methodology, Software, Formal analysis, Investigation, Visualization, Writing (original draft).\\ \textbf{Angélique Drémeau}: Conceptualization, Methodology, Supervision, Writing (review and editing).\\
\textbf{Arnaud Coatanhay}: Conceptualization, Methodology, Supervision, Writing (review and editing).}

\section{\chg{Conflict of interest statement}}
\chg{The authors declare no conflicts of interest.}

\section{\chg{Data availability statement}}
\chg{The data that support the findings of this study are available from the corresponding author upon reasonable request.}

\bibliographystyle{IEEEtran}
\bibliography{Biblio}

\end{document}